\documentclass[12pt]{article}
\usepackage{draft}

\begin{document}

\unitlength = .8mm

\begin{titlepage}

\begin{center}

\hfill \\
\hfill \\
\vskip 1cm

\title{Deformations with Maximal Supersymmetries
\\ Part 2: Off-shell Formulation}

\author{Chi-Ming Chang$^a$, Ying-Hsuan Lin$^a$, Yifan Wang$^b$, Xi Yin$^{a}$}

\address{$^a$Jefferson Physical Laboratory, Harvard University, \\
Cambridge, MA 02138 USA
\\
$^b$Center for Theoretical Physics, Massachusetts Institute of Technology, \\
Cambridge, MA 02139 USA}

\email{cmchang@physics.harvard.edu, yhlin@physics.harvard.edu, \\ yifanw@mit.edu,
xiyin@fas.harvard.edu}

\end{center}

\abstract{ Continuing our exploration of maximally supersymmetric gauge theories (MSYM) deformed by higher dimensional operators, in this paper we consider an off-shell approach based on pure spinor superspace and focus on constructing supersymmetric deformations beyond the first order. In particular, we give a construction of the Batalin-Vilkovisky action of an all-order non-Abelian Born-Infeld deformation of MSYM in the non-minimal pure spinor formalism. We also discuss subtleties in the integration over the pure spinor superspace and the relevance of Berkovits-Nekrasov regularization. }

\vfill

\end{titlepage}

\eject

\tableofcontents

\section{Introduction}

Methods of formulating maximally supersymmetric gauge theories with all 16 supersymmetries manifest have been developed,  in the on-shell formulation \cite{Siegel:1978yi} based on the algebra of super-gauge covariant derivatives and its deformations \cite{Movshev:2003ib,Movshev:2004aw,Movshev:2005ei,Movshev:2009ba}, and in the off-shell formulation based on pure spinor superspace by Cederwall and Karlsson \cite{Cederwall:2011vy,Cederwall:2013vba}, after the work of Berkovits \cite{Berkovits:2001rb}. 
The algebraic, on-shell approach was explored in our previous paper \cite{partone} to classify infinitesimal deformations that preserve 16 supersymmetries, while allowing the possibility of breaking either Lorentz or R-symmetry. In this approach, the problem of finding higher order deformations (or identifying the obstructions) can be formulated systematically as a cohomology problem. In practice, however, it was very difficult to compute the relevant obstruction classes and to verify their triviality.

In this paper, we adopt the off-shell approach based on pure spinor superspace.
This formalism was first developed in the context of superstring perturbation theory \cite{Nilsson:1985cm,Tonin:1991ii,Howe:1991mf,Howe:1991bx,Berkovits:2000fe,Berkovits:2000nn,Berkovits:2000ph,Berkovits:2001ue,Berkovits:2001us,Matone:2002ft,Berkovits:2002zk,Berkovits:2004px,Berkovits:2004tw,Berkovits:2005bt,Berkovits:2006vi}. It was known for some time that the standard two-derivative, undeformed, MSYM can be reformulated as a Chern-Simons-like theory in pure spinor superspace \cite{Siegel:1978yi, Berkovits:2001rb}, in close analogy with cubic open string field theory \cite{Witten:1985cc}. Although, it was not immediately obvious how to write down higher derivative deformations in this language. It was explained in \cite{Cederwall:2011vy} how the Born-Infeld deformation, to first order, can be constructed in the non-minimal pure spinor superspace formalism, and that the first order deformation in the Abelian case already gives a consistent action to all orders. We will develop this construction further, and show that the non-Abelian Born-Infeld deformation can be extended  {\it to all orders}. (See \cite{Bergshoeff:1986jm, Bergshoeff:2000cx, Bergshoeff:2001dc, Sevrin:2001ha,Cederwall:2001bt,Cederwall:2001td,Cederwall:2001dx,Collinucci:2002ac, Howe:2010nu,Bossard:2010pk} for previous works in the conventional component field formalism.) This is achieved in the BV formalism \cite{Batalin:1981jr,Fuster:2005eg}, where the question of finding higher order deformations of the action that solve the BV master equation is turned into a problem of showing the triviality of certain cohomology classes. We will see close analogies with the on-shell algebraic approach, and how introducing non-minimal pure spinor variables helps solving the problem. We also describe similar constructions in other examples, including the noncommutative deformation and the 5-form deformation (the latter in zero spacetime dimension only).

We note an important subtlety in dealing with the higher order deformations, as well as the construction of D-terms, in the non-minimal pure spinor formalism. Inversion of pure spinor variables is used in writing the descendant superfields, and the higher derivative terms in pure spinor superspace. This could potentially lead to divergences in the integration over the tip of the pure spinor cone. Such divergences do not seem to appear in our construction of F-term deformations, but this is not a priori obvious. We find it useful to consider a regularization introduced by Berkovits and Nekrasov \cite{Berkovits:2006vi}, which amounts to smearing the superfields in pure spinor superspace in a manner that preserves the BV master equation. This allows us to demonstrate the absence of divergences in simple examples. We suspect that it is relevant for the construction of  general D-terms in this formalism as well.

In section 2 we will review the pure spinor superspace and the descendant pure spinor superfields of \cite{Cederwall:2011vy}, and set up our notations and conventions. In section 3, we apply this formalism to the Born-Infeld deformation, and demonstrate that the construction can be extended to all orders in the deformation parameter, solving the BV master equation \cite{Batalin:1981jr,Fuster:2005eg} order by order. Other examples such as noncommutative deformations, and the 5-form deformation in the IKKT matrix model \cite{Ishibashi:1996xs}, are discussed in section 4. We introduce the Berkovits-Nekrasov regulator in the context of MSYM theories in section 5, and discuss their role in regularizing potential divergences in the pure spinor integral, and possibly the construction of D-terms. We conclude with some open questions in section 6.

\section{Pure spinor superspace}

In this section we review the construction of the action of maximally supersymmetric Yang-Mills theories based on pure spinor superspace. A first attempt at constructing an action based on the Yang-Mills superfield involves a Chern-Simons type functional defined by an integration over the ``minimal" pure spinor superspace. We will see that this action gives rise to the correct SYM equation of motion up to pure gauge terms, provided that a truncation on the superfield is implemented. The truncation condition breaks manifest supersymmetry, however. To fix the problem, one extends the superfield to one defined over the non-minimal pure spinor superfields  \cite{Cederwall:2011vy}. Instead of a classical gauge invariant action, in this formalism one find a Batalin-Vilkovisky action functional \cite{Batalin:1981jr,Fuster:2005eg,Cederwall:2011vy}. A conventional BRST invariant action may be obtained by imposing the Siegel gauge condition that effectively eliminates the BV anti-fields in the pure spinor superfield. Working with the BV action has the advantage that deformations of BRST transformations need not be introduced explicitly, but rather is determined via the BV anti-bracket. The problem of finding supersymmetric higher derivative deformations turns into the problem of constructing higher derivative terms that solve the BV master equation \cite{Cederwall:2011vy}. We will also see later that the closure of BV master equation order by order can be reformulated as a cohomology problem.

\subsection{Super-Yang-Mills theory and the pure spinor superfield}

Let us begin by considering the classical action of ${\cal N}=1$ SYM in 10 dimensions. The dimensional reduction to $d$ dimensional (undeformed) MSYM will be straightforward. Let $(x^m, \theta^\A)$ be superspace coordinates, $m=0,\cdots,9$, $\A=1,\cdots, 16$. The ordinary Yang-Mills superfield is written as $A_\A(x,\theta)$. The super-derivative is written as
\ie
& d_\A = {\partial\over\partial\theta^\A} - (\Gamma^m\theta)_\A {\partial\over\partial x^m}.
\fe
It obeys the anti-commutator
\ie\label{ddc}
\{ d_\A, d_\B \} = -2\Gamma^m_{\A\B} {\partial\over\partial x^m}.
\fe
Let $\lambda^\A$ be a pure spinor variable, namely it obeys the constraint
\ie
\lambda^\A \Gamma^m_{\A\B}\lambda^\B = 0.
\fe
The ordinary SYM equation of motion can be written in the form \cite{Siegel:1978yi}
\ie
\lambda^\A \lambda^\B (d_\A A_\B + A_\A A_\B) = 0.
\fe
If we write a pure spinor superfield $\Psi(x,\theta,\lambda)$ as $\lambda^\A A_\A(x,\theta)$, then the equation of motion can be put in the simple form
\ie\label{udeom}
Q\Psi + \Psi^2=0,
\fe
where\footnote{Although $Q$ is analogous to the BRST charge in the worldsheet formulation of pure spinor string theory, here in the context of spacetime gauge theory it is merely a differential and should not be confused with the BRST charge.}
\ie
& Q = \lambda^\A d_\A
\fe
is a nilpotent differential, namely $Q^2=0$, by virtue of (\ref{ddc}) and the pure spinor constraint on $\lambda$. Super-gauge transformations $\delta A_\A = d_\A \Omega + [A_\A, \Omega]$ can be expressed in terms of $\Psi$ as
\ie
\delta \Psi = Q\Omega + [\Psi,\Omega].
\fe
As an example, in the Abelian case, when $A_\A(x,\theta)$ obeys the equation of motion, there is a gauge in which $A_\A$ can be put in the form
\ie
A_\A(x,\theta) = {1\over 2}(\Gamma^m\theta)_\A a_m(x) + {1\over 12}(\theta\Gamma^{mnp}\theta) (\Gamma_{mnp})_{\A\B} \chi^\B(x)  + {1\over 16} (\theta \Gamma^{mnp}\theta) ( \Gamma_p\theta)_\A \partial_m a_{n}  + \cdots,
\fe
where $\cdots$ involves derivatives of $a_m$ and $\chi^\A$. $a_m(x)$ and $\chi^\A(x)$ are the component fields for the gauge boson and the gaugino.

So far the pure spinor superfield $\Psi$ is by definition linear in $\lambda^\A$, or in other words, it has ghost number 1, where the ghost number here simply counts the degree in $\lambda$ (the notion of ghost number will be extended later when we consider the non-minimal formalism). 
In writing a BRST invariant action or a BV action functional, it will be useful to relax the ghost number condition on $\Psi(x,\theta,\lambda)$, and allow for components of all degrees in $\lambda$:
\ie
\Psi(x,\theta,\lambda) = C(x,\theta) + \lambda^\A A_\A (x,\theta) + (\lambda\gamma^{mnpqr}\lambda) A^*_{mnpqr}(x,\theta) + \lambda^\A\lambda^\B \lambda^\C C^*_{\A\B\C}(x,\theta) + \cdots,
\fe
where $\cdots$ stands for terms that involve more than 3 powers of $\lambda$ (such terms will not play any role in the minimal formalism). Here $C(x,\theta)$ is the ordinary ghost superfield, $A^*_{mnpqr}$ and $C^*_{\A\B\C}$ are BV anti-fields. Note that the pure spinor constraint implies that $\lambda^{\A_1}\cdots\lambda^{\A_n}$ transforms under Lorentz group or $Spin(10)$ in a single irreducible representation of Dynkin label $[0000n]$.

A first attempt of writing a superspace action in connection with the Yang-Mills superfield equation is the following Chern-Simons-like action \cite{Berkovits:2001rb}
\ie\label{csa}
S = \int d^{10} x \,{\rm Tr} \left\langle {1\over 2}\Psi Q\Psi + {1\over 3}\Psi^3 \right\rangle.
\fe
Here ${\rm Tr}$ stands for the trace over the gauge index of $\Psi$. $\langle\cdots\rangle$ amounts to an integration over the minimal pure spinor superspace, with a peculiar choice of measure. It is {\it defined} to be nonzero only when evaluated on the $spin(10)$ singlet constructed out of $\lambda^3\theta^5$,
\ie
\big\langle (\lambda\Gamma^m\theta)(\lambda\Gamma^n\theta)(\lambda\Gamma^p\theta)(\theta\Gamma_{mnp}\theta) \big\rangle = 1,
\fe
and vanishes on any other monomials of the form $\lambda^k\theta^\ell$. This seemingly ad hoc definition has a natural explanation in the language of non-minimal pure spinor superspace, which will be reviewed in section \ref{nonmin}. Note that while this measure has the property that $\langle Q(\cdots)\rangle$ is a total derivative, $\langle\partial_{\theta^\A}(\cdots)\rangle$ generally is not a total derivative.

If we restrict $\Psi$ to its ghost number 1 component, of the form $\lambda^\A A_\A(x,\theta)$, then the expression (\ref{csa}) reduces to a gauge invariant functional of $A_\A(x,\theta)$. 
%
The problem is that varying it with respect to $A_\A(x,\theta)$ does not quite reproduce the equation of motion $Q\Psi+\Psi^2=0$. For instance all terms involving 5 or more $\theta$'s in $\Psi$ drop out of the action functional. Such terms will end up as pure gauge, but still (\ref{csa}) is not quite the correct action in the conventional sense. This problem will be resolved in the non-minimal pure spinor formalism, where infinitely many more auxiliary fields are introduced.

\subsection{Reducing to component fields}

It is instructive nonetheless to inspect explicitly the functional (\ref{csa}) restricted to $\Psi = \lambda^\A A_\A(x,\theta)$. Since the resulting functional is gauge invariant under $\delta A_\A = d_\A \Omega + [A_\A, \Omega]$, let us first restrict the form of $A_\A(x,\theta)$ using such a gauge transformation. 
For simplicity, we will illustrate with the example zero-dimensional MSYM (also known in its component field form as the IKKT matrix model), where $A_\A$ is a function of $\theta$ only, and the gauge transformation takes the form 
\ie
\delta A_\A(\theta) = {\partial\over \partial\theta^\A}\Omega + [A_\A, \Omega].
\fe
We can remove $A_\A(0)$ with a linear gauge parameter in $\theta$. Let $A_\A^{(n)}$ be the degree $n$ component of $A_\A$ in $\theta$. Now the minimal action can be written as
\ie
S &= \left\langle \lambda^\A\lambda^\B\lambda^\C {\rm Tr} \left( A_\A^{(1)} {\partial\over\partial \theta^\B} \left.A_\C^{(5)}\right|_{[00030]}+A_\A^{(2)} {\partial\over\partial \theta^\B} \left.A_\C^{(4)}\right|_{[10020]} \right.\right.
\\
&~~~~~\left.\left. + {1\over 2} A_\A^{(3)} {\partial\over\partial \theta^\B} A_\C^{(3)} + A_\A^{(1)} A_\B^{(1)} A_\C^{(3)}+ A_\A^{(1)} A_\B^{(2)} A_\C^{(2)} \right) \right\rangle.
\fe
Note that here we only retain dependence on the representation component $[00030]$ in $A^{(5)}$ (which contains $\Lambda^5[00001]=[00030]\oplus [11010]$), and the component $[10020]$ in $A^{(4)}$ (which contains $\Lambda^4[00001] = [02000]\oplus [10020]$).
Varying with respect to $A^{(5)}$ and $A^{(4)}$ gives the following equations
\ie
& \lambda^\A \lambda^\B \partial_\A A_\B^{(1)}=0 ~\Rightarrow~ A_\A^{(1)} = (\Gamma^m \theta)_\A a_m,
\\
& \lambda^\A\lambda^\B\partial_\A A_\B^{(2)} = 0~\Rightarrow~ A_\A^{(2)} = (\theta\Gamma_{mnp}\theta) (\Gamma^{mnp})_{\A\B} \chi^\B.
\fe
These conditions remove some of the gauge redundancy in $A_\A^{(1)}$ and $A_\A^{(2)}$ while retaining the physical degrees of freedom, the gauge boson $a_m$ and the gaugino $\chi^\A$. Varying with respect to $A^{(3)}$ gives 
\ie
\lambda^\A\lambda^\B \left( \partial_\A A_\B^{(3)} + A_\A^{(1)} A_\B^{(1)} \right) = 0,
\fe
which is the precisely the degree 2 component of the equation $Q\Psi+\Psi^2=0$ with $A^{(0)}$ set to zero. Here we have used the fact that $\theta^3$ contains only 1 irreducible representation of $spin(10)$, namely $\Lambda^3[00001]= [01010]$. We have also used the fact that in the minimal pure spinor superspace integral we can integrate by parts on $Q$ (but not on $\partial_{\theta^\A}$ by itself).  Next, if we vary $A^{(2)}$ and $A^{(1)}$, we obtain
\ie
&\lambda^\A\lambda^\B \left( \partial_\A \left.A_\B^{(4)}\right|_{[10020]} + \{A_\A^{(1)}, A_\B^{(2)}\} \right) = 0,
\\
&\lambda^\A\lambda^\B \left( \partial_\A \left.A_\B^{(5)}\right|_{[00030]} + \left.\{ A_\A^{(1)}, A_\B^{(3)}\}\right|_{[10020]} + \left.A_\A^{(2)} A_\B^{(2)}\right|_{[10020]} \right) = 0.
\fe
The first equation is the correct restriction of $Q\Psi+\Psi^2=0$ to degree 3, keeping only the $[10020]$ representation component of $A^{(4)}$. The second equation is the restriction of $Q\Psi+\Psi^2=0$ to degree 4 and the representation $[10020]$, keeping only the $[00030]$ component of $A^{(5)}$. Note that the degree 4, $[02000]$ component of the equation of motion is missing here. However, this component of $Q\Psi+\Psi^2=0$ would have only involved the $[11010]$ component of $A^{(5)}$, which has dropped out the minimal superspace action altogether. As a result, we do get the correct equation of motion for $A^{(5)}|_{[00030]}$.

To summarize, once we have fixed on the gauge condition $A^{(0)}=0$, the only components of $A_\A(\theta)$ that appears in the minimal pure spinor action $S$ are given by
\ie\label{resa}
A_\A = \left.A_\A^{(1)}\right|_{[00001]} + \left.A_\A^{(2)}\right|_{[00100]} + \left.A_\A^{(3)}\right|_{[01010]} + \left.A_\A^{(4)}\right|_{[10020]} + \left.A_\A^{(5)}\right|_{[00030]}.
\fe
The resulting equations by varying $S$ with respect to these components are precisely the restriction of the equation $Q\Psi+\Psi^2=0$ to the relevant components. These equations then give the correct SYM equations for $a_m$ and $\chi^\A$, in zero dimension.

The price to pay, if we make the restriction (\ref{resa}), is that the super-gauge invariance is no longer manifest, since the gauge variation $\delta A_\A$ generally does not maintain the form (\ref{resa}). This is expected, since we cannot implement 16 off-shell supersymmetries with finitely many auxiliary fields \cite{galperin2007harmonic}. The way to cure this problem is to introduce the non-minimal pure spinor variables, which allows for writing down the superspace action with a conventional measure, and no restriction of the form (\ref{resa}) on the pure spinor superfield will be needed.

\subsection{The non-minimal pure spinor superspace}
\label{nonmin}

In order to write down higher order terms in the pure spinor superfield, one needs some way of taking derivative with respect to the pure spinor variable $\lambda$, as the naive $\partial/\partial\lambda^\A$ is generally not well defined due to the constraints. This is achieved through the non-minimal pure spinor variables, as was first introduced in the context of pure spinor string theory \cite{Berkovits:2005bt}. We must pay a hefty price however: infinitely many more auxiliary fields are introduced, and generally we will need to work in the BV formalism \cite{Cederwall:2009ez}.

One introduces a new ``conjugate" pure spinor $\bar\lambda_\A$, of the opposite chirality as $\lambda^\A$, that obeys $\bar\lambda \Gamma^m\bar\lambda=0$. It will be also necessary to introduce a Grassmannian variable $r_\A$ that obeys $\bar\lambda\gamma_m r=0$. $r_\A$ can be identified with the differential $d\bar\lambda_\A$, and we will sometimes use this notation when it does not cause confusion. The differential $Q$ will be modified to
\ie
Q = \lambda^\A d_\A + r_\A {\partial\over \partial \bar\lambda_\A}.
\fe
Note that the combination $r_\A \partial_{\bar\lambda_\A}$ annihilates $\bar\lambda \gamma^m \bar\lambda$ due to the constraint on $r$, and thus is well defined. This is also clear if we think of $r_\A \partial_{\bar\lambda_\A} = d\bar\lambda_\A \partial_{\bar\lambda_\A}$ as taking exterior derivative on $\bar\lambda$.

Now we will extend the pure spinor Yang-Mills superfield $\Psi(x,\theta,\lambda)$ to one that depends on $\bar\lambda, r$ also, $\Psi(x,\theta,\lambda,\bar\lambda,r)$. This introduces infinitely many more auxiliary fields, but does not change the number of physical degrees of freedom because the cohomology of $Q$ in the $(\bar\lambda,r)$ sector is trivial. The superspace integration will take the form
\ie
\int d^{16}\theta [d\lambda] [d\bar\lambda] [dr],
\fe
where the $spin(10)$ invariant measure factors $[d\lambda], [d\bar\lambda], [dr]$ are defined as
\ie
& [d\lambda] \lambda^\A \lambda^\B \lambda^\C = (\epsilon \bar T)^{\A\B\C}_{\A_1\cdots\A_{11}}d\lambda^{\A_1}\cdots d\lambda^{\A_{11}},
\\
& [d\bar\lambda] \bar\lambda_\A \bar\lambda_\B \bar\lambda_\C = (\epsilon T)_{\A\B\C}^{\A_1\cdots\A_{11}}d\bar\lambda_{\A_1}\cdots d\bar\lambda_{\A_{11}},
\\
& [dr] = (\epsilon \bar T)^{\A\B\C}_{\A_1\cdots\A_{11}} \bar\lambda_\A \bar\lambda_\B \bar\lambda_\C {\partial\over \partial r_{\A_1}}\cdots {\partial\over \partial r_{\A_{11}}}.
\fe
Here $T$ is the $spin(10)$ invariant tensor defined by
\ie
(\lambda\Gamma^m\theta)(\lambda\Gamma^n\theta)(\lambda\Gamma^p\theta)(\theta\Gamma_{mnp}\theta)
= T_{\A\B\C\A_1\cdots\A_5} \lambda^\A \lambda^\B\lambda^\C \theta^{\A_1}\cdots\theta^{\A_5},
\fe
and $\epsilon T$ its contraction with the 16-dimensional anti-symmetric tensor. $\bar T$ is the same tensor with chiral and anti-chiral spinors exchanged. In performing the integration of $(\lambda, \bar\lambda)$ over the pure spinor superspace, $\bar\lambda_\A$ will be regarded as the complex conjugate variable of $\lambda^\A$. Note that we could have also simplified our notation by identifying $r_\A$ with $d\bar\lambda_\A$ and write the integration measure as
\ie
\int d^{16}\theta [d\lambda],
\fe
while the $d^{11}\bar\lambda$ factor will be supplied from the integrand which is now regarded as a differential form in $\bar\lambda_\A$ rather than a function of $r_\A$.

The superfield will be regarded as an analytic function in the pure spinor variables $\lambda, \bar\lambda$. In order for the integration over the pure spinor space to converge as $\lambda,\bar\lambda\to \infty$, one multiplies the integration measure with a regulator of the form
\ie
\exp(-\zeta\{Q, \Lambda\}).
\fe
It is crucial that such a regulator formally differs from 1 by a $Q$-exact expression, so as to ensure that $Q$-exact integrands integrate to zero. A convenient choice is
\ie
\Lambda = \bar\lambda_\A \theta^\A,~~~~ \{Q,\Lambda\} = \bar\lambda_\A\lambda^\A + r_\A\theta^\A.
\fe
Note that the BV action constructed by integrating with this regulator, as a functional of $\Psi$, will generally depend on $\zeta$, since the integrand isn't $Q$-closed. Note that the dependence on $\zeta$ would drop out if we restrict to the part of integrand of homogeneous degree 3 in $\lambda$ and $r$. 

Now the superspace SYM action is written as
\ie\label{undeformedaction}
S = \int d^{10} x d^{16}\theta [d\lambda] [d\bar\lambda] [dr] e^{-\zeta(\bar\lambda\lambda + r\theta)}\, {\rm Tr} \left( {1\over 2}\Psi Q\Psi + {1\over 3}\Psi^3 \right).
\fe
If we restrict $\Psi$ to be independent of $\bar\lambda, r$, then the $(\theta,\lambda,\bar\lambda,r)$ measure factor may be replaced by
\ie
\int d^{16}\theta [d\lambda]  e^{-\zeta\bar\lambda\lambda} {(-\zeta d\bar\lambda\theta)^{11}\over 11!}
\fe
which is nonzero only when evaluated with the integrand $(\lambda^3\theta^5)$, giving
\ie
&\int d^{16}\theta [d\lambda]  e^{-\zeta\bar\lambda\lambda} {(-\zeta d\bar\lambda\theta)^{11}\over 11!}
\lambda^\A\lambda^\B\lambda^\C = 
\int d^{16}\theta (\epsilon T)^{\A\B\C}_{\A_1\cdots\A_{11}}\theta^{\A_1}\cdots \theta^{\A_{11}}
\\
&= T^{\A\B\C \A_1\cdots\A_5} \left.{\partial\over\partial\theta^{\A_1}}\cdots {\partial\over\partial\theta^{\A_5}}\right|_{\theta=0}.
\fe
This is the minimal pure spinor superspace measure we have seen in the previous subsection.

Let us denote collectively $Z=(\lambda, \theta, \bar\lambda, r)$, and the regularized non-minimal superspace integration measure as $[dZ] = d^{10}x d^{16}\theta [d\lambda]  [d\bar\lambda] [dr] e^{-\zeta\{Q,\Lambda\}}$. Given two functionals $F,G$ of $\Psi$, one may define a Batalin-Vilkovisky antibracket \cite{Cederwall:2009ez} by
\ie
(F, G) = - {\rm Tr}\int [dZ] {\delta F\over \delta\Psi(Z)} {\delta G\over \delta\Psi(Z)}.
\fe
The sign convention has to do with the fact that our measure factor $[dZ]$ is odd.
The extension of the nilpotency of BRST transformation in the BV formalism is the (classical) BV master equation
\ie
(S, S) = 0.
\fe


\subsection{Descendant pure spinor superfields}

A key ingredient introduced by \cite{Cederwall:2011vy} is the construction of descendant superfields from $\Psi$ by acting with certain linear differential operators. The first few descending operators are $\hat A_\A$, $\hat A_m$, $\hat \chi^\A$, $\hat F_{mn}$, $\widehat \eta_n^\A$. They obey the descending relations
\ie\label{descending}
& [Q, \hat A_\A] = -d_\A - 2(\Gamma^m\lambda)_\A \hat A_m.
\\
& \{ Q, \hat A_m \} = \partial_m - \lambda\Gamma_m \hat\chi,
\\
& [Q,\hat \chi^\A] = -{1\over 2}(\Gamma^{mn}\lambda)^\A \hat F_{mn},
\\
& \{Q, \hat F_{mn} \} = 2 \lambda\Gamma_{[m} \hat\eta_{n]}.
\fe
Explicitly, they are given by
\ie\label{amans}
&\hat A_\A =- (\lambda\bar\lambda)^{-1} \left[ {1\over 4} \bar\lambda_\A N + {1\over 8} (\Gamma^{mn}\bar\lambda)_\A N_{mn} \right],
\\
& \hat A_m = -{1\over 4} (\lambda\bar\lambda)^{-1} (\bar\lambda\Gamma_m d) + {1\over 32} (\lambda\bar\lambda)^{-2} {(\bar\lambda\Gamma_{mnp}r)} N^{np},
\\
& \hat\chi^\A = {1\over 2}(\lambda\bar\lambda)^{-1} (\Gamma^m \bar\lambda)^\A \partial_m - {1\over 192}(\lambda\bar\lambda)^{-2} (\bar\lambda\Gamma^{mnp}r) (\Gamma_{mnp}d)^\A
- {1\over 64} (\lambda\bar\lambda)^{-3} (\Gamma_m\bar\lambda)^\A (r\Gamma^{mnp} r) N_{np},
\\
& \hat F_{mn} = {1\over 8}(\lambda\bar\lambda)^{-2}(\bar\lambda\Gamma_{mn}{}^p r)\partial_p
+ {1\over 32} (\lambda\bar\lambda)^{-3} (r \Gamma_{mnp}r)(\bar\lambda\Gamma^p d) - {1\over 256} (\lambda\bar\lambda)^{-4} (\bar\lambda \Gamma_{mnp} r) (r\Gamma^{pqr} r) N_{qr},
\fe
where $N=\lambda^\A{\partial\over\partial\lambda^\A}$ and $N_{mn}=\lambda^\A(\Gamma_{mn})_{\A}{}^\B{\partial\over\partial\lambda^\B}$. It will be convenient to introduce an operator $\Delta_m$,
\ie
\Delta_m \equiv \partial_m + {1\over 4}(\lambda\bar\lambda)^{-1} (r\Gamma_m d) - {1\over 32} (\lambda\bar\lambda)^{-2} (r\Gamma_{mnp}r)N^{np}.
\fe
$\Delta_m$ is analogous to $\partial_m$ but has a nontrivial commutator with $Q$,
\ie{}
[Q, \Delta_m] = {1\over 2}(\lambda\bar\lambda)^{-1} (r\Gamma_m \Gamma^n \lambda) \Delta_n.
\fe
This property will be useful later in constructing deformations of the BV action. The descending operators $\hat\chi^\A, \hat F_{mn}$ and $\hat \eta_m^\A$ are related to $\Delta_m$ by
\ie
& \hat\chi^\A = {1\over 2}(\lambda\bar\lambda)^{-1} (\Gamma^m\bar\lambda)^\A \Delta_m,
~~~~~ ~~ \hat F_{mn} = {1\over 8} (\lambda\bar\lambda)^{-2} (\bar\lambda\Gamma_{mn}{}^p r) \Delta_p,
\\
& \hat\eta^\A_m = -{1\over 16}(\lambda\bar\lambda)^{-3} (\Gamma_n\bar\lambda)^\A (r\Gamma_m{}^{np} r)\Delta_p = -{1\over 32} (\lambda\bar\lambda)^{-2} (r\Gamma_{mnp}r)(\Gamma^{np}\hat\chi)^\A.
\fe
A useful fact is that all of $\hat\chi^\A$ and $\hat F_{mn}$ anti-commute or commute with one another.\footnote{
The LiE package~\cite{lie} and the decomposition of tensor products of $r$ and $\bar\lambda$ into irreps of $SO(10)$ listed in~\cite{Cederwall:2011vy} are useful in verifying the relations among the descendent operators.
}
Note that $\lambda\Gamma^m\hat\chi$ and $\lambda\Gamma^n\hat\chi$ do {\it not} commute,\footnote{There appears to be an incorrect statement regarding this in \cite{Cederwall:2011vy}.} though they would commute when their indices are contracted with $\hat F_{mn}$ or $\hat F_{mp}\cdots \hat F_{nq}\cdots$. The following relations are also useful:
\ie
& \hat\chi\Gamma_m\hat\chi = 0,~~~~ (\lambda\Gamma^m\hat\chi) \hat F_{mn} = \hat F_{mn} (\lambda\Gamma^m\hat\chi) = 0,
\\
& [\Delta_m, \Delta_n] = 0,~~~~ [\Delta_m, \lambda\bar\lambda]=0.
\fe

\subsection{Recovering the on-shell Yang-Mills superfields}
\label{recmin}

We will later construct deformations of the undeformed MSYM action in the sense of Batalin-Vilkovisky in non-minimal pure spinor superspace, generally of the form
\ie
S = \int [dZ] {\rm Tr} \left( {1\over 2} \Psi Q\Psi + {1\over 3} \Psi^3 \right)  + \sum_{n=1}^\infty \epsilon^n S^{(n)}[\Psi],
\fe
where $S^{(n)}[\Psi]$ will be an integral over the non-minimal superspace of a function of linear descendant fields $\hat \chi\Psi$, $\hat F\Psi$, etc. The BV master equation will be solved order by order in the deformation parameter $\epsilon$. Since $\Psi(x,\theta,\lambda,\bar\lambda,r)$ now contains infinitely many auxiliary fields, here we would like to describe how to recover a deformed equation of motion for an ordinary Yang-Mills superfield $A_\A(x,\theta)$.

We will consider an analogous expansion of a ghost number 1 superfield $\Psi$ in $\epsilon$,
\ie
\Psi = \Psi_0(x,\theta,\lambda) + \sum_{n=1}^\infty \epsilon^n \Psi_n(x,\theta,\lambda,\bar\lambda, r),~~~~\Psi_0 = \lambda^\A A_\A(x,\theta).
\fe
Suppose $\Psi_0$ solves the equation $Q\Psi_0+\Psi_0^2=0$ of the undeformed MSYM theory. We would like to construct a nearby solution of the deformed theory. To first order in $\epsilon$, the equation to solve is
\ie
Q\Psi_1 + \{\Psi_0, \Psi_1\} + \left.{\delta S^{(1)}\over \delta\Psi}\right|_{\Psi_0} = 0.
\fe
The key is to show that $\delta S^{(1)}/\delta \Psi$ evaluated on $\Psi_0$ can be put in the form
\ie
 \left.{\delta S^{(1)}\over \delta\Psi}\right|_{\Psi_0} = {\cal E}_1[\Psi_0] + Q\Lambda + \{\Psi_0, \Lambda\},
\fe
where the term ${\cal E}_1[\Psi_0]$ involves {\it only the minimal variables}, and $\Lambda$ is a function of $\Psi_0$ and its derivatives that generally involves non-minimal variables. If we can do this, then we would have recovered the first order deformation of the equation on minimal superfield $\Psi_{min}(x,\theta,\lambda)$ as
\ie
Q\Psi_{min} + \Psi_{min}^2 + \epsilon\, {\cal E}_1[\Psi_{min}] = {\cal O}(\epsilon^2).
\fe
Here $\Psi$ simply differs from $\Psi_{min}$ by $\epsilon \Lambda$.

In practice, we can construct ${\cal E}_1[\Psi_0]$ from $S^{(1)}$ roughly by replacing the linear descendant fields with the minimal descendant superfields. To illustrate this, let us consider the example of Abelian Born-Infeld theory, with
\ie\label{sone}
\left.{\delta S^{(1)}\over \delta\Psi}\right|_{\Psi_0} &= (\lambda\Gamma^m\hat\chi \Psi_0) (\lambda\Gamma^n\hat\chi \Psi_0) (\hat F_{mn}\Psi_0)
\\
&= -{1\over 2}(\lambda\bar\lambda)^{-2}(r\Gamma^{mnp}\bar\lambda) (\Delta_m\Psi_0 )(\Delta_n\Psi_0)( \Delta_p\Psi_0).
\\
\fe
The expected ${\cal E}_1[\Psi_0]$ is 
\ie\label{eone}
{\cal E}_1[\Psi_0] = (\lambda\Gamma^m\chi)(\lambda\Gamma^n\chi)F_{mn}, 
\fe
where $\chi^\A$ and $F_{mn}$ are the {\it minimal} descendant superfields, related to $A_\A(x,\theta)$ via
\ie
& A_m = -{1\over 16}\Gamma_m^{\A\B} d_\A A_\B,~~~~ \chi^\A = -{1\over 10}\Gamma_m^{\A\B}(d_\B A^m - \partial_m A_\B),
~~~~ F_{mn} = \partial_m A_n - \partial_n A_m.
\fe
For $\lambda^\A d_\A$-closed $\Psi_0(x,\theta)$, both (\ref{sone}) and (\ref{eone}) are $Q$-closed. Generally, the existence of ${\cal E}_1[\Psi_0]$ is a consequence of the statement that the non-minimal variables $\bar\lambda, r$ do not introduce new $Q$-cohomology. What we need to see here is that (\ref{sone}) and (\ref{eone}) differ by a $Q$-exact term, thus verifying in particular that the off-shell deformation is a nontrivial one.

We can write
\ie
& (\lambda\Gamma^m\chi)(\lambda\Gamma^n\chi)F_{mn}
= (\lambda\Gamma^m\chi)(\lambda\Gamma^n\chi)\left[ - {1\over 2}(\lambda\bar\lambda)^{-1}(r\Gamma_{mn}\chi) +{1\over 2}(\lambda\bar\lambda)^{-2}(\lambda r)(\bar\lambda\Gamma_{mn}\chi) \right]
\\
&~~~~ + Q\left[ {1\over 2}(\lambda\bar\lambda)^{-1} (\lambda\Gamma^m\chi)(\lambda\Gamma^n\chi)(\bar\lambda\Gamma_{mn}\chi) \right]
\\
&=-{1\over 2} (\lambda\bar\lambda)^{-2} (r\Gamma^{mnp}\bar\lambda) (\lambda\Gamma_m\chi)(\lambda\Gamma_n\chi)(\lambda\Gamma_p\chi)
 + Q\left[ {1\over 2}(\lambda\bar\lambda)^{-1} (\lambda\Gamma^m\chi)(\lambda\Gamma^n\chi)(\bar\lambda\Gamma_{mn}\chi) \right].
\fe
This is now very close to (\ref{sone}), but there is still a little difference between $\Delta_m\Psi_0$ and $\lambda\Gamma_m\chi$. We have
\ie
\Delta_m\Psi_0 &= \lambda^\A \partial_m A_\A + {1\over 4}(\lambda\bar\lambda)^{-1} (r\Gamma_m d) (\lambda A) - {1\over 32}(\lambda\bar\lambda)^{-2} (r\Gamma_{mij} r)(\lambda\Gamma^{ij} A)
\\
&= \lambda\Gamma_m\chi + Q\left[ A_m + {1\over 4}(\lambda\bar\lambda)^{-1} (\bar\lambda\Gamma_m d) (\lambda A)  - {1\over 32}(\lambda\bar\lambda)^{-2} (\bar\lambda\Gamma_{mij} r)(\lambda\Gamma^{ij} A)\right] 
\\
&~~~~ + ({\rm stuff~that~vanishes~upon~contraction~with~}r\Gamma^{mnp}\bar\lambda)
\\
&= \lambda\Gamma_m\chi + Q\left[ {1\over 2}  (\lambda\bar\lambda)^{-1} (\lambda\Gamma_m \Gamma^k \bar\lambda)A_k 
-{1\over 8} (\lambda\bar\lambda)^{-2} (\lambda\Gamma_m \Gamma_k r) (\bar\lambda\Gamma^k A)
\right] 
\\
&~~~~ + ({\rm stuff~that~vanishes~upon~contraction~with~}r\Gamma^{mnp}\bar\lambda).
\fe
Now we can put (\ref{sone}) in the form
\ie\label{sones}
\left.{\delta S^{(1)}\over \delta\Psi}\right|_{\Psi_0} &= -{1\over 2}(\lambda\bar\lambda)^{-2}(r\Gamma^{mnp}\bar\lambda) (\lambda\Gamma_m \chi' )(\lambda\Gamma_n \chi' )( \lambda\Gamma_p \chi' ),
\fe
where 
\ie
\chi'^\A = \chi^\A + Q \left[ {1\over 2}  (\lambda\bar\lambda)^{-1} (\Gamma^k \bar\lambda)^\A A_k 
-{1\over 8} (\lambda\bar\lambda)^{-2} (\Gamma_k r)^\A (\bar\lambda\Gamma^k A) \right]
\fe
Using the identity
\ie
Q \left[  (\lambda\bar\lambda)^{-1} (\lambda\Gamma^m)_{[\A} (\lambda\Gamma^n)_\B (\bar\lambda\Gamma_{mn})_{\C]} \right] = (\lambda\bar\lambda)^{-2} (r\Gamma^{mnp}\bar\lambda)
(\lambda\Gamma_m)_\A (\lambda\Gamma_n)_\B (\lambda\Gamma_p)_\C,
\fe
which in particular implies that the RHS commutes with $Q$, we see that (\ref{sones}) is indeed equal to (\ref{eone}) up to $Q$-exact terms.


\section{The Born-Infeld deformation}

A primary example of interest in this paper is the Born-Infeld deformation of MSYM theory. At the infinitesimal level, this is an F-term deformation of the Lagrangian by a dimension 8 operator. While this deformation is expected to preserve all 16 supersymmetries, in the usual component field formalism the Lagrangian deformation is only invariant under supersymmetries up to terms proportional to the equation of motion, which must be compensated by deformation of the supersymmetry transformations. Such a procedure generally requires adding terms to all orders in the deformation parameter, and there could be potential obstructions in finding higher order terms. The Abelian Born-Infeld theory to all orders in the deformation parameter (a.k.a. $\A'^2$ in the context of string theory) was first constructed in \cite{Aganagic:1996nn} by gauge fixing a kappa symmetric D-brane action. It seemed difficult to generalize this approach to the non-Abelian case.

In the conventional component field formalism, the second order Born-Infeld deformation was constructed in \cite{Howe:2010nu}. Using pure spinor superspace, an all-order Abelian Born-Infeld deformation was constructed in \cite{Cederwall:2011vy}. It was not clear whether the action of \cite{Cederwall:2011vy} upon integrating out auxiliary fields would coincide with the construction from the super D-brane action. A priori they could differ by D-terms. The objective of this section is to extend the construction of \cite{Cederwall:2011vy} in the non-Abelian case to all orders in the deformation parameter. In principle this also gives a solution to the on-shell deformation problem, considered in Part 1 of the paper \cite{partone}.

\subsection{The first order deformation}

Let us begin by recalling the construction of the infinitesimal Born-Infeld deformation in BV formalism based on non-minimal pure spinor superspace \cite{Cederwall:2011vy}. 
This is described by a quartic deformation of the MSYM action,\footnote{We shall use $S_n$ to denote the part of the BV action $S$ with degree $n$ in $\Psi$.} 
\ie
\label{bif}
S_4 ={ \epsilon\over 4} \int [dZ] {\rm Tr} \left[ \Psi \circ (\lambda\Gamma^m\hat\chi\Psi) \circ (\lambda\Gamma^n\hat\chi\Psi) \circ (\hat F_{mn}\Psi) \right],
\fe
where $\circ$ denotes the symmetric product.  Variation with respect to $\Psi$ corrects the equation of motion to
\ie
Q\Psi+\Psi^2 + \epsilon(\lambda\Gamma^m\hat\chi\Psi) \circ (\lambda\Gamma^n\hat\chi\Psi) \circ (\hat F_{mn}\Psi) = 0,
\fe
which is cohomologically equivalent to the on-shell Born-Infeld deformation in terms of minimal pure spinor superfields, in the sense explained in section \ref{recmin}. In showing this, one integrates by part with respect to the differential operators $(\lambda\Gamma^m\hat\chi)$ and $\hat F_{mn}$, making use of the identities $(\lambda\Gamma^{m}\hat\chi)(\lambda\Gamma^{n }\hat\chi) (\bar\lambda \Gamma_{mnp}r)=0$ and $(\lambda\Gamma^m\hat\chi)\hat F_{mn} = 0$. Note that despite the presence of the regulator $e^{-\zeta (\lambda\bar\lambda + r\theta)}$ in $[dZ]$, $\hat\chi^\A$ and $\hat F_{mn}$ in fact commute with this regulator. Note also that while $\hat\chi^\A$ does not commute with $\lambda^\B$, they satisfy $\lambda\Gamma^m\hat\chi = \hat\chi\Gamma^m\lambda$. 
%
This infinitesimal deformation actually does not depend on the value of the parameter $\zeta$ in the regulator. If we vary $\zeta$ in \eqref{bif}, we obtain a term that can be written as $(S_2+S_3, G)$, where
\ie
G= \int [dZ]\bar\lambda^\A d_\A {\rm Tr} \left[ \Psi \circ (\lambda\Gamma^m\hat\chi\Psi) \circ (\lambda\Gamma^n\hat\chi\Psi) \circ (\hat F_{mn}\Psi) \right].
\fe
The integrand inside $(S_2+S_3, G)$ is proportional to the undeformed equation of motion \eqref{udeom}; hence, it can be  absorbed by field redefinition of $\Psi$. 

To see that the action of the form $S_2+S_3+S_4$ obeys BV master equation up to order $\epsilon$, one needs to show that $(S_2, S_4)=0$ and $(S_3,S_4)=0$. The manipulations needed to verify these relations will be useful for the extension to higher order deformations later, and so let us recall how this is done. Firstly, we have
\ie\label{stsf}
& (S_2, S_4) =
{\epsilon \over 4} \int [dZ] {\rm Tr}\left[ Q\Psi\circ (\lambda\Gamma^m\hat\chi \Psi)\circ (\lambda\Gamma^n\hat\chi \Psi)\circ (\hat F_{mn}\Psi) 
- \Psi\circ (\lambda\Gamma^m\hat\chi Q\Psi)\circ (\lambda\Gamma^n\hat\chi \Psi)\circ (\hat F_{mn}\Psi )
\right.
\\
&~~~\left. + \Psi\circ (\lambda\Gamma^m\hat\chi \Psi)\circ (\lambda\Gamma^n\hat\chi Q\Psi)\circ (\hat F_{mn}\Psi) - \Psi\circ (\lambda\Gamma^m\hat\chi \Psi)\circ (\lambda\Gamma^n\hat\chi \Psi)\circ (\hat F_{mn} Q\Psi)  \right].
\fe
Using the fact that $Q$ commutes with $\lambda\Gamma^m\hat\chi$, $\{Q, \hat F_{mn} \} = 2\lambda \Gamma_{[m} \hat\eta_{n]}$, and the basic pure spinor identity $(\lambda\Gamma^m)_\A(\lambda\Gamma_m)_\B=0$,
we see that the integrand is $Q$-exact and thus $(S_2, S_4)=0$.

To see the vanishing of $(S_3, S_4)$, it is useful to use the identity
$
\hat F_{mn} = -{1\over 4}(\lambda\bar\lambda)^{-1} (r\Gamma_{mn}\hat\chi),
$
and rewrite $S_4$ as
\ie
S_4 = {\epsilon\over 4}  \int [dZ](\lambda\bar\lambda)^{-1} (\Gamma^m\lambda)_{[\A}(\Gamma^n\lambda)_\B(\Gamma_{mn}r)_{\C]}  {\rm Tr}\left( \Psi\circ \hat\chi^\A \Psi\circ \hat\chi^\B \Psi\circ \hat\chi^\C \Psi \right).
\fe
Further using $\hat\chi^\A = {1\over 2}(\lambda\bar\lambda)^{-1}(\Gamma^m\bar\lambda)^\A\Delta_m$, we can write
\ie
S_4 &= -{\epsilon\over 16} \int [dZ] (\lambda\bar\lambda)^{-4} (\lambda\Gamma_m\Gamma^i\bar\lambda)(\lambda\Gamma_n\Gamma^j\bar\lambda)(r\Gamma^{mnk}\bar\lambda){\rm Tr}\left[ \Psi\circ \Delta_i \Psi\circ \Delta_j \Psi\circ \Delta_k \Psi \right]
\\
&= - {\epsilon\over 8} \int [dZ] (\lambda\bar\lambda)^{-2}(r\Gamma^{ijk}\bar\lambda){\rm Tr}\left( \Psi \Delta_i \Psi \Delta_j \Psi \Delta_k \Psi \right).
\fe
In going to the second line, we used the pure spinor constraints on $\bar\lambda_\A$ and $r_\A$, which in particular implies $(\lambda\Gamma_m\Gamma^i\bar\lambda)(\lambda\Gamma_n\Gamma^j\bar\lambda)(r\Gamma^{mnk}\bar\lambda)= 4 (\lambda\bar\lambda)^2(r\Gamma^{ijk}\bar\lambda)$.
In the last trace we can replace the symmetric product by ordinary, due to the symmetry on $[ijk]$.
The BV bracket with $S_3$ is computed as
\ie
(S_3, S_4) &= - {\epsilon\over 8} \int [dZ] (\lambda\bar\lambda)^{-2}(r\Gamma^{ijk}\bar\lambda){\rm Tr}\left( \Psi^2 \Delta_i \Psi \Delta_j \Psi \Delta_k \Psi - \Psi\Delta_i \Psi^2 \Delta_j \Psi \Delta_k \Psi 
\right.
\\
&~~~~\left. + \Psi\Delta_i \Psi \Delta_j \Psi^2 \Delta_k \Psi - \Psi \Delta_i \Psi \Delta_j \Psi \Delta_k \Psi^2 \right)
\\
&= {\epsilon\over 8} \int [dZ] (\lambda\bar\lambda)^{-2}(r\Gamma^{ijk}\bar\lambda){\rm Tr}\left( \Psi^2 \circ\Delta_i \Psi\circ \Delta_j \Psi\circ \Delta_k \Psi
\right).
\fe
%
We've chosen to rewrite the last line in terms of a symmetrized product once again, for later convenience.
The representation in terms of the $\Delta$'s is particularly useful due to the properties $[\Delta_i,\Delta_j]=0$, $[\Delta_i,\lambda\bar\lambda]=0$, and $\int [dZ] \Delta_i(\cdots)=0$ which allows for integration by parts on $\Delta_i$. Repeatedly applying integration by parts and cyclicity of the trace, as well as the anti-symmetry on $[ijk]$, we can make the following replacement on the integrand
\ie
& {\rm Tr}\left( \Psi^2 \circ\Delta_i \Psi\circ \Delta_j \Psi\circ \Delta_k \Psi \right) 
\to {3\over 2}{\rm Tr}\left(\{\Psi, \Delta_i\Psi\} \circ \Psi\circ \Delta_j \Psi\circ \Delta_k \Psi
\right)
\\
&\to {3\over 2} {\rm Tr}\left(\Delta_i\Psi \circ \{\Psi, \Psi\circ \Delta_j \Psi\circ \Delta_k \Psi\}
\right)
\\
&\to  3{\rm Tr}\left(\Delta_i\Psi \circ \Psi^2\circ \Delta_j \Psi\circ \Delta_k \Psi
\right) +3{\rm Tr}\left(\Delta_i\Psi\circ \Psi\circ \Delta_j \Psi\circ \Delta_k \Psi^2
\right)
\\
&\to 6{\rm Tr}\left( \Psi^2 \circ\Delta_i \Psi\circ \Delta_j \Psi\circ \Delta_k \Psi \right) .
\fe
This shows that indeed $(S_3, S_4)=0$, thus completing the verification that the Born-Infeld deformation (\ref{bif}) solves the BV master equation at order $\epsilon$. Now at order $\epsilon^2$, there is a potentially non-vanishing contribution to the BV master equation,
\ie\label{sfsf}
(S_4, S_4) = - 3 \epsilon^2 \int [dZ] (\lambda\bar\lambda)^{-4} (r\Gamma^{ijk}\bar\lambda)(r\Gamma^{mnp}\bar\lambda){\rm Tr}(\Delta_i\Psi\Delta_j\Psi\Delta_k\Psi\Delta_m\Psi\Delta_n\Psi\Delta_p\Psi).
\fe
Note that the combination $r^2\bar\lambda^2$ appearing in the prefactor of the integrand can only transform in the representation $[00120]\oplus [01011]$ of $spin(10)$, due to the pure spinor constraints on $r$ and $\bar\lambda$. In the case of {\it Abelian} gauge theory, $\Delta_i\Psi\cdots\Delta_p\Psi$ lives in the 6th anti-symmetric tensor representation, or $[00011]$. It cannot form a singlet by contracting with $r^2\bar\lambda^2$, and hence (\ref{sfsf}) vanishes in the Abelian theory. It does not vanish in the non-Abelian case, and a second order deformation of the action must be introduced to cancel this term in the BV master equation. This will be analyzed next.

\subsection{Non-Abelian Born-Infeld deformation at the second order}

Let us now consider $(S_4, S_4)$ in the non-Abelian theory. 
Using Baker-Campbell-Hausdorff formula, we can write the integrand in $(S_4, S_4)$ as
\ie\label{sfour}
& (\lambda\bar\lambda)^{-4}(r\Gamma^{ijk}\bar\lambda)(r\Gamma^{mnp}\bar\lambda){\rm Tr}\bigg[\big(\Delta_i\Psi\circ \Delta_j\Psi\circ\Delta_k\Psi\big) \big(\Delta_m\Psi\circ \Delta_n\Psi\circ \Delta_p\Psi\big) \bigg]
\\
&= (\lambda\bar\lambda)^{-4}(r\Gamma^{ijk}\bar\lambda)(r\Gamma^{mnp}\bar\lambda){\rm Tr}\bigg[ \Delta_i\Psi\circ \Delta_j\Psi\circ\Delta_k\Psi\circ \Delta_m\Psi\circ \Delta_n\Psi\circ \Delta_p\Psi
\\
&~~- {9\over 2} \Delta_i\Psi\circ \Delta_m \Psi\circ \{\Delta_j\Psi, \Delta_n \Psi\} \circ \{\Delta_k\Psi, \Delta_p \Psi\} + 3 \Delta_i\Psi\circ \Delta_m\Psi\circ \Delta_n\Psi\circ [\Delta_j\Psi, \{ \Delta_k\Psi, \Delta_p\Psi \}]
\\
&~~ + {3\over 4} \{ \Delta_i\Psi, \Delta_m\Psi \}\circ \{\Delta_j\Psi, [\Delta_n\Psi, \{ \Delta_p\Psi, \Delta_k\Psi \}]\}
-{1\over 4} [\Delta_i\Psi, \{\Delta_j\Psi, \Delta_m\Psi\}]\circ [\Delta_n\Psi,\{\Delta_p\Psi,\Delta_k\Psi\}]
\\
&~~ - {1\over 5} \Delta_i\Psi\circ \big[ \Delta_j\Psi,\{\Delta_m\Psi, [\Delta_n\Psi, \{ \Delta_p\Psi, \Delta_k\Psi \}]\}\big] - {3\over 5} \Delta_i\Psi\circ \big[ \Delta_m\Psi,\{\Delta_j\Psi, [\Delta_n\Psi, \{ \Delta_p\Psi, \Delta_k\Psi \}]\}\big] \bigg].
\\
&= {3\over 2} (\lambda\bar\lambda)^{-4} (r\Gamma^{ijk}\bar\lambda)(r\Gamma^{mnp}\bar\lambda){\rm Tr}\bigg[ \Delta_i\Psi\circ \Delta_m \Psi\circ \{\Delta_j\Psi, \Delta_n \Psi\} \circ \{\Delta_k\Psi, \Delta_p \Psi\} 
\\
&~~~~~~~~ - {1\over 5} [\{ \Delta_i\Psi, \Delta_m\Psi \},\Delta_j\Psi]\circ [\Delta_n\Psi, \{ \Delta_p\Psi, \Delta_k\Psi \}]\} \bigg].
\fe
In above we used the fact that the term appearing in the second line is zero, for the same reason as in the Abelian case, and simplified the rest using cyclicity of the trace.
The resulting expression is nonzero, and we would like to cancel it by an order $\epsilon^2$ deformation of the action. A priori, one may try to cancel it with either $(S_2, S_6)$, by adding to $S$ some sextic term $S_6$, or with $(S_3, S_5)$, by adding a quintic term $S_5$. It is easy to see that this cannot be done using $S_6$. The reason is that we would have to construct $S_6$ by taking $r\bar\lambda^3$ contracted with the trace of a product of 6 $\Delta\Psi$'s. However, $r\bar\lambda^3$ consists of the representations $[00040]\oplus [00120]$ of $spin(10)$, and neither appear in the (unsymmetrized) 6-fold tensor power of the vector representation, and so no such singlet exist as a candidate for $S_6$.

On the other hand, it is possible to cancel $(S_4, S_4)$ by introducing a quintic term $S_5$, such that $(S_4,S_4) + 2(S_3,S_5)=0$. Let us first consider the term on the RHS of (\ref{sfour}) that involves a 4-fold symmetric product. Firstly, we have the identity
\ie\label{zzt}
& \int [dZ]  (\lambda\bar\lambda)^{-4} (r\Gamma^{ijk}\bar\lambda)(r\Gamma^{mnp}\bar\lambda){\rm Tr}
\bigg[ \Delta_i\Psi\circ \Delta_m \Psi\circ \{\Delta_j\Psi, \Delta_n \Psi\} \circ \{\Delta_k\Psi, \Delta_p \Psi \}\bigg]
\\
&= \int [dZ]  (\lambda\bar\lambda)^{-4} (r\Gamma^{ijk}\bar\lambda)(r\Gamma^{mnp}\bar\lambda){\rm Tr}
\bigg[
3 \Psi^2\circ \Delta_i \Psi \circ \Delta_m \Psi \circ \{ \Delta_n\Delta_k \Psi, \Delta_j\Delta_p \Psi \}
\\
&~~~~  - 8 \Psi^2  \circ \Delta_m \Psi \circ \Delta_j\Delta_n \Psi \circ \{ \Delta_k\Delta_p \Psi, \Delta_i \Psi \}
\bigg].
\fe
In this manipulation we used  integration by parts on the $\Delta_i$'s (recall that $\Delta_i$ also commutes with $\lambda\bar\lambda$), the cyclicity of the trace, and the symmetry on the indices $[ijk][mnp]$. It is now easy to write down an $S_5$ such that $(S_3, S_5)$ can be used to cancel the term appearing in (\ref{zzt}). We must at the same time make sure that $S_5$ has vanishing BV anti-bracket with $S_2$. This can be achieved by rewriting expressions involving $\Delta_i$'s in terms of $\lambda\Gamma^m\hat\chi$ and $(\lambda\Gamma^{mn})^\A\hat F_{mn}$,
\ie\label{QchiF}
& \lambda\Gamma^m\hat\chi = {1\over 2}(\lambda\bar\lambda)^{-1} (\lambda\Gamma^m \Gamma^n\bar\lambda) \Delta_n,
\\
& (\lambda\Gamma^{mn})^\A \hat F_{mn} = {1\over 8}(\lambda\bar\lambda)^{-2} (\lambda\Gamma^{mn})^\A (\bar\lambda\Gamma_{mn}{}^p r) \Delta_p,
\fe
both of which commute with $Q$. One can verify that the quintic term that can be used to cancel the RHS of (\ref{sfour}) is 
\ie
S_5&=-{3\over 4}\int [dZ](\lambda\bar\lambda)^{-4} (r\Gamma^{ijk}\bar\lambda)(r\Gamma^{mnp}\bar\lambda){\rm Tr} \Big[\Psi\circ \Delta_i \Psi \circ \Delta_m \Psi \circ \{ \Delta_n\Delta_k \Psi, \Delta_j\Delta_p \Psi \}
\\
&~~~~~~~~-2\Psi \circ \Delta_m \Psi \circ \Delta_j\Delta_n \Psi \circ \{ \Delta_k\Delta_p \Psi, \Delta_i \Psi \}-{1\over 5}\Psi\circ \left[\{\Delta_j\Psi,\Delta_i \Delta_m\Psi\} , \{ \Delta_p\Psi, \Delta_n\Delta_k\Psi \}\right]\Big]
\\
&=-48\int [dZ]\, {\rm Tr} \bigg[\Psi\circ (\lambda\Gamma_i\hat\chi) \Psi \circ (\lambda\Gamma_m\hat\chi) \Psi \circ \{ (\lambda\Gamma_n\hat\chi)\hat F_{ij} \Psi, (\lambda\Gamma_j\hat\chi)\hat F_{mn} \Psi \}
\\
&~~~~~~~~~~~~~~~~~~~~-2\Psi \circ (\lambda\Gamma_m\hat\chi) \Psi \circ (\lambda\Gamma_j\hat\chi)(\lambda\Gamma_n\hat\chi) \Psi \circ \{\hat F_{ij}\hat F_{mn} \Psi,(\lambda\Gamma_i\hat\chi)\Psi \}
\\
&~~~~~~~~~~~~~~~~~~~~-{1\over 5}\Psi\circ \left[\{(\lambda\Gamma_i\hat\chi)\Psi,(\lambda\Gamma_j\hat\chi)\hat F_{mn}\Psi\} , \{ (\lambda\Gamma_m\hat\chi)\Psi, (\lambda\Gamma_n\hat\chi)\hat F_{ij}\Psi \}\right]\bigg].
\fe

The way we could solve for an $S_5$ with the property $(S_3, S_5) = -{1\over 2}(S_4, S_4)$ is no accident. The essential point is that $(S_4, S_4)$ is closed with respect to $(S_3, \,\cdot\,)$, and the operation $(S_3, \,\cdot\,)$, which is nilpotent and can be regarded as a coboundary operator on the space of functionals of $\Psi$, has trivial cohomology in this case. We will demonstrate this more generally in the next subsection.

\subsection{No obstruction to all order}

In this section, we will prove the existence of an all-order formal deformation that solves the BV master equation, whose first order term in the deformation parameter $\epsilon$ is $S_4$. Firstly, note that the BV anti-bracket satisfies Jacobi identity
\ie\label{abc}
(A,(B,C))=((A,B),C)+(-1)^{|A||B|}(B,(A,C)),
\fe
where $A,B,C$ are functionals of $\Psi$. We define an odd differential $\widehat\delta A\equiv (S_3,A)$ on functionals of $\Psi$, that obeys ${\widehat\delta}\,^2A=0$ because of the Jacobi identity and $(S_3,S_3)=0$.\footnote{It is important here that the BV anti-bracket is even and $S_3$ has odd degree by convention of (\ref{abc}) (which is shifted by 1 from the usual convention). Of course, it is also easy to verify directly that $\widehat\delta\,^2=0$. } The BV anti-bracket of two $\widehat\delta$-closed functionals is $\widehat\delta$-closed, and the BV anti-bracket of a $\widehat\delta$-closed functional with a $\widehat\delta$-exact functional is $\widehat\delta$-exact. So in other words, the BV anti-bracket defines a cup product on the cohomology of $\widehat\delta $. Note that in fact, the cohomology of $\widehat\delta$ is defined already on the traces of products of derivatives of $\Psi$, without the need for integration over pure spinor superspace.

Now consider the vector space ${\cal V}$ spanned by functionals constructed by taking $\hat F_{mn}$'s and $(\lambda\Gamma^m\hat\chi)$'s acting on $\Psi$, with all vector indices on the $(\lambda\Gamma^m\hat\chi)$'s contracted with those of the $\hat F$'s, traced and then integrated over the pure spinor superspace. A typical functional of this type looks like
\ie\label{Fchi}
\int[dZ]\tr\left[\cdots \hat F_{mn}\hat F_{pq}\Psi(\lambda\Gamma^n\hat\chi)\Psi\cdots(\lambda\Gamma^q\hat\chi)\Psi(\lambda\Gamma^m\hat\chi)\Psi\cdots(\lambda\Gamma^p\hat\chi)\Psi\cdots\right].
\fe
The virtue of this construction is that, due to \eqref{QchiF}, such a functional has vanishing BV anti-brackets with $S_2$. Furthermore, the BV anti-bracket of two such functionals remains in ${\cal V}$. 

The action of $\widehat\delta$ on such functionals, on the other hand, is simplified if we express $\hat F_{mn}$ and $\lambda\Gamma^m\hat\chi$ in terms of $\Delta_m$, using
\ie
&\hat F_{mn}=-{1\over 8}(\lambda\bar\lambda)^{-2}(r\Gamma_{mn}\Gamma^p\bar\lambda)\Delta_p,
~~~~{\rm and}~~~~
\lambda\Gamma^m\hat\chi={1\over 2}(\lambda\bar\lambda)^{-1}(\lambda\Gamma^m\Gamma^n\bar\lambda)\Delta_n.
\fe
By construction, here the vector index on $\lambda\Gamma^m\hat\chi$ is always contracted with an index on an $\hat F_{mn}$, and so the $\lambda\Gamma^m\hat\chi$ will always appears in the combination $(r\Gamma_{mn}\Gamma_p\bar\lambda)\lambda\Gamma^m\hat\chi$, which can be simplified as
\ie
& (r\Gamma_{mn}\Gamma_p\bar\lambda)\lambda\Gamma^m\hat\chi={1\over 2}(\lambda\bar\lambda)^{-1}(r\Gamma_{mn}\Gamma_p\bar\lambda)(\lambda\Gamma^m\Gamma^q\bar\lambda)\Delta_q
\\
&={1\over 2}(\lambda\bar\lambda)^{-1}(r\Gamma_{np}\Gamma_m\bar\lambda)(\lambda\Gamma^m\Gamma^q\bar\lambda)\Delta_q
=(r\Gamma_{np}\Gamma_m\bar\lambda)\Delta^m=(r\Gamma_{mn}\Gamma_p\bar\lambda)\Delta^m.
\fe
In other words, on any of the functionals in ${\cal V}$, we can replace $\lambda\Gamma_m\hat\chi$ by $\Delta_m$. Next, because $\Delta_m$ commutes with $\lambda\bar\lambda$ (and trivially commutes with $r,\bar\lambda$),
after expressing $\hat F$ in terms of $\Delta$, we can move all explicit factors involving $r,\lambda,\bar\lambda$ outside the $\Delta$'s and outside the trace. A functional in ${\cal V}$ can thus be rewritten as a linear combination of the terms
\ie\label{fnl}
\int [dZ]T^{i_1\cdots i_{3n}}\tr\left[\left(\Delta_{i_{w(1)}}\cdots\Delta_{i_{w(p_1)}}\Psi\right)\left(\Delta_{i_{w(p_1+1)}}\cdots\Delta_{i_{w(p_2)}}\Psi\right)\cdots \left(\Delta_{i_{w(p_{m-1})+1}}\cdots\Delta_{i_{w(p_m)}}\Psi\right)\right],
\fe
where $0\le p_1\le\cdots\le p_m= 3n$,\footnote{If $p_{\ell-1}$ and $p_{\ell}$ coincide then by convention there is no $\Delta$ acting on $\Psi$ in the $\ell$-th factor.} $w$ is an element of the permutation group $S_{3n}$ on $\{1,\cdots,3n\}$, and $T^{i_1\cdots i_{3n}}=(\lambda\bar\lambda)^{-2n}(r\Gamma^{i_1i_2i_3}\bar\lambda)\cdots(r\Gamma^{i_{3n-2}i_{3n-1}i_{3n}}\bar\lambda)$.  Since $T^{i_1\cdots i_{3n}}$ commutes with $\Delta_k$, we are free to integrate by part on the $\Delta$'s.

The tensor $T^{(n)}\sim T^{i_1\cdots i_{3n}}$ transforms in the overlap between the representation content of $r^n \bar\lambda^n$ (as listed in the table of \cite{Cederwall:2011vy}) and $\Lambda^n[00100]$. We list these representations below:\footnote{Interestingly, the absence of $T^{(11)}$ ensures that we do not have a term with $(\lambda\bar\lambda)^{-11}$ pole in the integrand, that would come with 11 powers of $r$. If such a term were present, it would lead to a log divergence in the $(\lambda,\bar\lambda)$ integral, making the action ill defined.}
\ie
&T^{(1)}\in [00100],
\\
& T^{(2)}\in [01011],
\\
& T^{(3)}\in[02020] \oplus [10022],
\\
& T^{(4)}\in [00033]\oplus [11031],
\\
& T^{(5)} \in [01042] \oplus [20140],
\\
& T^{(6)} \in [10151]\oplus [30060],
\\
& T^{(7)}\in [00260]\oplus [20071],
\\
& T^{(8)}\in [10180],
\\
& T^{(9)}\in [0,1,0,10,0],
\\
& T^{(10)}\in [0,0,0,12,0],
\\
& T^{(11)}\equiv 0.
\fe
The structures in $S_{n+3}$ that we will encounter at the $n$-th order in the deformation parameter are of the schematic form $T^{(n)} {\rm tr} \Delta^{3n}\Psi^{n+3}$.

The cohomology of $\widehat\delta$ on ${\cal V}$ is equivalent to a certain invariant cyclic cohomology. Let $t_i$ be a set of commutative variables, $i=1,\cdots,10$ (they can be thought of as dual variables to the $\Delta_i$'s that act on a single $\Psi$ inside the trace). Let ${\bf h}=\langle t_i\rangle$ be the Abelian Lie algebra generated by commuting variables $t_i$ (i.e. the linear vector space spanned by the $t_i$'s), and $A=U({\bf h})=\mathbb{C}[t_i]$ its universal enveloping algebra. Let $C^k = {\rm Hom} (\otimes^{k+1} A,\mathbb{C})$ be the Hochschild cochains. The cyclic complex $C_\lambda^k$ is obtained by taking the part of $C^k$ that is invariant under the $\mathbb{Z}_{k+1}$ that shifts the $k+1$ arguments with sign, namely
\ie
C_\lambda^k = \{\varphi\in C^k: \varphi(a_k,a_0,\cdots, a_{k-1}) = (-1)^k \varphi(a_0,a_1,\cdots,a_k) \}.
\fe
The differential $\delta: C_\lambda^k \to C_\lambda^{k+1}$ defined by
\ie
(\delta \varphi )(a_0,\cdots, a_{k+1}) = \sum_{i=0}^k (-)^i \varphi(a_0,\cdots,a_i a_{i+1},\cdots, a_{k+1})
+(-)^{k+1}\varphi(a_{k+1}a_0,\cdots,a_k)
\fe
is nilpotent. The cohomology of $\delta$ at $C_\lambda^k$ defines the cyclic cohomology ${\rm HC}^k(A)$. Next, consider the complex ${\cal C}^{\ell,k}=\Lambda^\ell {\bf h}^*\otimes C_\lambda^k$ with the chain map $d:\Lambda^\ell {\bf h}^*\otimes C_\lambda^k\to \Lambda^{\ell-1} {\bf h}^*\otimes C_\lambda^k$ defined by
\ie
& d(\eta_1\wedge\cdots\wedge \eta_{\ell}\otimes\varphi)(a_0,\cdots,a_k) 
\\
&= \sum_{i=1}^\ell \sum_{j=0}^k (-1)^{i-1} \eta_1\wedge\cdots\wedge \eta_{i-1}\wedge \eta_{i+1}\wedge\cdots \wedge \eta_\ell\otimes\varphi(a_0,\cdots,a_{j-1}, \eta_i (a_j), a_{j+1}, \cdots,a_k).
\fe
Here ${\bf h}^*$ is the dual Lie algebra of ${\bf h}$, generated by $\partial/\partial t^i$, and $\eta(a)$ is defined as the derivative map for $a\in A=\mathbb{C}[t_i]$, $\eta\in{\bf h}^*$.
In other words, with $C_\lambda^k$ viewed as an ${\bf h}^*$-module as above, the cohomology of the complex $\Lambda^\ell {\bf h}^* \otimes C_\lambda^k$ with respect to the differential $d$ defines the Lie algebra homology ${\rm H}_\ell({\bf h}^*,C_\lambda^k)$. It is easy to see that $d\delta = \delta d$.

Now $\delta$ induces a map on ${\rm H}_\ell({\bf h}^*, C_\lambda^k)$,
\ie
\delta_*: {\rm H}_\ell({\bf h}^*, C_\lambda^k) \to {\rm H}_\ell({\bf h}^*, C_\lambda^{k+1}).
\fe
The cohomology of $\delta_*$ on ${\rm H}_0({\bf h}^*, C^k_\lambda)$ defines the invariant cyclic cohomology ${\rm HC}^k_{{\bf h}^*}(A)$. The cohomology of $\widehat\delta$ on the space of functionals ${\cal V}$ at the $n$-order (with $n+3$ $\Psi$'s) lies in ${\rm HC}^{n+2}_{{\bf h}^*}(A)$; they correspond to the components that transform under $spin(10)$ according to the representations of $T^{(n)}$.




The ordinary cyclic {cohomology} ${\rm HC}^k(A)$ can be computed using a homology version of Grothendieck's algebraic de Rham complex \cite{LodayQuillen}, 
\ie\label{hcc}
& {\rm HC}^k(A) \simeq \left.{\rm Ker}\,d^*\right|_{(\Omega^k)^*} \oplus {\rm H}_{k-2,dR}(A)\oplus {\rm H}_{k-4,dR}(A)\oplus\cdots
\fe
where $(\Omega^k)^*$ is the space of de Rham $k$-currents in the $t_i$'s. $d^*$ is the transpose of de Rham differential. $H_{*,dR}(A)$ is the algebraic de Rham homology of $A$, which coincides with the ordinary de Rham homology on $Spec(A)=V\simeq \mathbb{C}^{10}$, defined in terms of the codifferential on polyvector fields. 
For odd $k$, ${\rm HC}^k(A)\simeq {\rm Ker}\,d^*|_{(\Omega^k)^*}$ is the dual vector space of the cokernal of $d:\Omega^{k-1}\to \Omega^k$. In this case due to the triviality of de Rham homology we can also identify it as the dual space of $d\Omega^k\subset \Omega^{k+1}$. These classes are in correspondence with the (unintegrated) traces of derivatives of $\Psi$'s of the form\footnote{There is a canonical pairing between integrands of the form (\ref{opdd}) and algebraic de Rham differential forms in $d\Omega^k$,
\ie\label{pairing}
\left<f(\Delta_i){\rm Tr}(\Delta_{j_1}\Psi\circ\cdots\circ\Delta_{j_{k+1}}\Psi)|\partial_{k}g(t_i)dt_{k}\wedge dt_{j_1}\wedge\cdots \wedge dt_{i_{k}}\right>=\delta^{i_1}_{[j_1}\cdots\delta^{i_k}_{j_k}f(\partial_i)\partial_{j_{k+1}]}g(t_i). 
\fe
Note that expressions of the form (\ref{opdd}) are not all independent: for intance, $\Delta_{[k}{\rm Tr}(\Delta_{j_1}\Psi\circ\cdots\circ\Delta_{j_{k+1}]}\Psi)\equiv 0$. This is precisely consistent with the pairing (\ref{pairing}). Therefore, we can identify the set of operators (\ref{opdd}) with $ (d\Omega_{k})^*\simeq {\rm Ker}(d^*)|_{(\Omega_k)^*}$.}
\ie\label{opdd}
\Delta_{i_1}\cdots\Delta_{i_m} {\rm Tr} \left( \Delta_{j_1}\Psi\circ\cdots\circ \Delta_{j_{k+1}}\Psi \right).
\fe
This statement is familiar in the context of counting BPS operators \cite{Chang:2013fba}. But since all of these term will end up giving total derivatives, they will not be relevant in the invariant cyclic cohomology of interest here.
For even $k$, there is an additional part of ${\rm HC}^k(A)$ coming from ${\rm H}_{0,dR}(A)\simeq \mathbb{C}$. This corresponds to the element ${\rm Tr}\Psi^{k+1}$, which is nonzero only for even $k$.


Now for the invariant cyclic cohomology ${\rm HC}^k_{{\bf h}^*}(A)$, there is an analogous relation with the invariant de Rham homology \cite{2002math......7118K} (with respect to the action of ${\bf h}^*$, by translation on the affine space in this case),
\ie
& {\rm HC}_{{\bf h}^*}^k(A) \simeq \left.{\rm Ker}\,d^*\right|_{(\Omega^k_{{\bf h}^*})^*} \oplus {\rm H}_{k-2,dR}^{{\bf h}^*}(A)\oplus {\rm H}_{k-4,dR}^{{\bf h}^*}(A)\oplus\cdots
\fe
The invariant de Rham homology on the RHS are simply represented by constant de Rham currents, i.e.
\ie
{\rm H}_{k-2\ell,dR}^{{\bf h}^*}(A) \simeq \Lambda^{k-2\ell} V.
\fe
Similarly, $\left.{\rm Ker}\,d^*\right|_{(\Omega^k_{{\bf h}^*})^*}\simeq \Lambda^k V$. So we conclude that
\ie
& {\rm HC}_{{\bf h}^*}^k(A) \simeq \bigoplus_{\ell\geq 0} \Lambda^{k-2\ell}V.
\fe
These class have clear interpretations in terms of the functionals of $\Psi$, of degree $n+3=k+1$. The $\Lambda^k V$ consists of $\widehat\delta$-closed integrals of the form
\ie
\int [dZ] {\rm Tr} \Big( \Psi\circ \Delta_{i_1}\Psi\circ\cdots\circ \Delta_{i_{n+2}}\Psi  \Big).
\fe
The remaining $\Lambda^{k-2\ell}V$ for $\ell\geq 1$ are represented by integrals of the form
\ie
\int [dZ] {\rm Tr} \Big( \Psi^{2\ell+1}\circ \Delta_{i_1}\Psi\circ\cdots\circ \Delta_{i_{n+2-2\ell}}\Psi +\cdots \Big),
\fe
where the $\cdots$ stands for terms of the same degree in $\Delta$ and $\Psi$ but with different orderings in the trace. The key point in conclusion is that the only nontrivial $\widehat\delta$-cohomology classes are represented by functionals in ${\cal V}$ that involve fewer $\Delta$'s than $\Psi$'s. This is enough to prove the absence of obstruction in solving the BV master equation for the Born-Infeld deformation to all orders.

Now we show that there is an all-order formal deformation of the form 
\ie
S=S_2+S_3+\sum\limits_{n=1}^\infty S_{n+3},
\fe
where $S_4$ is the first order non-Abelian Born-Infeld deformation (\ref{bif}), and $S_5,S_6,\cdots$ are functionals in ${\cal V}$. $S_{n+3}$ is of order $\epsilon^n$. We prove this by induction. Suppose $S_5,S_6,\cdots, S_M$ are all functionals of the form ${\cal V}$, and solve the BV master equation up to order $\epsilon^M$. Namely,
\ie\label{lower}
(S_3,S_n)=-{1\over 2}\sum^{n-1}_{i=4}(S_i,S_{n+3-i}),~~~~n=5,6,\cdots,M.
\fe
The order $\epsilon^{M+1}$ term in the BV master equation takes the form
\ie\label{dSM1}
\widehat\delta S_{M+1}=(S_3,S_{M+1})=-{1\over 2}\sum^{M}_{n=4}(S_n,S_{M+4-n}).
\fe
The RHS of this equation is a functional of the type ${\cal V}$, of degree $3M-3$ in $\Delta$ and $M+2$ in $\Psi$, and by (\ref{lower}) one sees that the RHS of \eqref{dSM1} is a $\widehat\delta$-closed. Namely,
\ie
\Big(S_3,\sum^{M}_{n=4}(S_n,S_{M+4-n})\Big)&=2\sum^{M}_{n=5}\big((S_3,S_n),S_{M+4-n}\big)
\\
&=-\sum^{M}_{n=5}\sum^{n-1}_{i=4}\big((S_i,S_{n+3-i}),S_{M+4-n}\big)=0.
\fe
In above the Jacobi identity on the BV anti-bracket was used repeatedly.
We have seen that cohomology of $\widehat\delta$ on ${\cal V}$ of such degrees in $\Delta$ and $\Psi$ is trivial. This means that the RHS of (\ref{dSM1}) is $\widehat\delta$-exact, and a solution for $S_{M+1}$ of the type ${\cal V}$ exists.

\section{Other examples}

In this section, we consider two examples of deformations that preserve maximal supersymmetries, but break either Lorentz invariance (noncommutative deformation) or R-symmetry (5-form deformation).

\subsection{Noncommutative deformation}

In every spacetime dimension $d$ between 0 and 10, besides the Born-Infeld deformation, there is only one class of maximally supersymmetric single trace F-term deformations that preserve the $Spin(10-d)$ R-symmetry. This is the noncommutative deformation of MSYM \cite{Seiberg:1999vs}. As was well known, it can be implemented by replacing the product of fields in MSYM action \eqref{undeformedaction} with a noncommutative associative $\star$-product, defined by
\ie
f(x)\star g(x)=f(x)\exp\left(\epsilon~\omega^{ij}\overleftarrow\partial_i\overrightarrow\partial_j\right)g(x),
\fe
where $\omega^{ij}$ is a constant 2-form (more precisely, a Poisson structure). We will fix $\omega^{ij}$ and think of the coefficient $\epsilon$ as an expansion parameter. Cyclicity of the trace is maintained up to total derivatives. Consequently the noncommutative-deformed action still solves the BV master equation. Expanding the deformed action in $\epsilon$ to the first order, we obtain the undeformed action $S_2+S_3$ plus
\ie
S_3'={2\over 3}\epsilon \int[dZ]~{\rm Tr}\Big(\omega^{ij}\Psi\partial_i\Psi\partial_j\Psi\Big).
\fe
An alternative and equivalent way to write the first order deformation in $\epsilon$ is\footnote{Written this way, the Born-Infeld deformation looks like a noncommutative deformation with the field strength $F_{ij}$ replacing the non-commutativity parameter $\omega_{ij}$. Though, of course, such a naive replacement would have resulted in a non-associative star product.}
\ie
S''_3+S''_4={2\over 3}\epsilon \int[dZ]~{\rm Tr}\Big(\omega^{ij}\Psi(\lambda\Gamma_i\hat\chi)\Psi(\lambda\Gamma_j\hat\chi)\Psi\Big)-{1\over 6}\epsilon  \int [dZ] \omega^{ij} {\rm Tr}\Big(\Psi^3\circ \hat F_{ij}\Psi\Big)
\fe
This differs from $S_3'$ by a term that can be removed by field redefinition at the first order. Namely, their difference is $(S_2+S_3)$-exact:
\ie\label{S3S3p}
S_3'-(S_3''+S_4'')=(S_2+S_3,G),
\fe
where $G$ is the functional
\ie
G [\Psi] =   {2\over 3}\epsilon\int [dZ] \omega^{ij} {\rm Tr}\left(\Psi \circ\hat A_i \Psi \circ (\partial_j+(\lambda\Gamma_j\hat\chi))\Psi\right).
\fe
To see this, we can compute the BV anti-brackets of $G$ with $S_2,S_3$ as
\ie
&(S_2,G)= {2\over 3}\epsilon \int [dZ] \omega^{ij} {\rm Tr}\left(\Psi\circ \{Q,\hat A_i\} \Psi \circ (\partial_j+(\lambda\Gamma_j\hat\chi)) \Psi\right),
\\
&(S_3,G)= {1\over 6}\epsilon  \int [dZ] \omega^{ij} {\rm Tr}\Big(\Psi^3\circ\hat F_{ij}\Psi\Big).
\fe
(\ref{S3S3p}) follows from the descending relation $\{Q, \hat A_i\} = \partial_i - \lambda\Gamma_i\hat\chi$.
Now the RHS of \eqref{S3S3p} is an integral whose integrand is proportional to the undeformed equation of motion. Therefore, the deformations by $S_3'$ and by $S_3''+S_4''$ are equivalent up to a field redefinition, modulo ${\cal O}(\epsilon^2)$ terms.

\subsection{The 5-form deformation}

An F-term deformation that is not an R-symmetry singlet exists in zero dimensional MSYM (IKKT matrix model), transforming in the self-dual 5-form representation of the $Spin(10)$ R-symmetry.\footnote{There are other R-symmetry breaking F-term deformations in general $d$ dimensions, that transform in the symmetric traceless tensor representation of $Spin(10-d)$. They may be viewed as a generalization of the Born-Infeld deformation. We will not discussion their off-shell constructions here.} This arises in the world volume theory of multi-D-instantons probing the $AdS_5\times S^5$ background of type IIB string theory, when viewed as a deformation of flat spacetime. The first order deformation of the action is given by
\ie\label{ffo}
 S_3'+S_4'=&\epsilon  \int [dZ]\omega^{\A\B} {\rm Tr}\Big[\Psi ((\Gamma^m\lambda)_\A\hat A_m\Psi)  ((\Gamma^n\lambda)_\B\hat A_n \Psi)\Big]
 \\
 &+{1\over 16}\epsilon  \int [dZ] \omega^{\A\B}{\rm Tr}\Big[\Psi^3([\hat A_\A,[Q,\hat A_\B]]\Psi) \Big],
\fe
where
\ie
\omega^{\A\B} \equiv \omega^{pqrst} (\Gamma_{pqrst})^{\A\B}.
\fe
Using $[Q, \hat A_\A] = - d_\A - 2(\Gamma^m\lambda)_\A \hat A_m$, it is easy to verify that
\ie
(S_2, S_3') = (S_3, S_3')+(S_2,S_4') = (S_3, S_4') = 0,
\fe
and so the BV master equation is obeyed at first order in $\epsilon$.

One may attempt to extend this deformation to all-order, by representing it as a noncommutative deformation in the superspace, with the Poisson structure given by $\omega^{\A\B}$. Namely, we replace the ordinary product in the undeformed action \eqref{undeformedaction} by a noncommutative $\star$-product defined as
\ie
f(\theta)\star g(\theta)=f(\theta)\exp\left(\epsilon~\omega^{\A\B}\overleftarrow d_\A\overrightarrow d_\B \right)g(\theta).
\fe
This is only well-defined on $(0|16)$ superspace, because in higher spacetime dimensions the superderivatives $d_\A$'s do not commute with one another.
Expanding the action to first order in $\epsilon$, one has
\ie\label{5form}
S_3'' = {2\over 3}\epsilon  \int [dZ] {\rm Tr}\left(\omega^{\A\B}\Psi (d_\A\Psi)  (d_\B \Psi)\right).
\fe
This amounts to replacing $(\lambda\Gamma^m)_\A\hat A_m$ in $S_3'$ by $d_\A$.
However, such a construction appears problematic because $d_\A$ does not commute with the regulator $\exp(-\zeta(\lambda\bar\lambda+r\theta))$, and so we would not be able to integrate by parts on $d_\A$. Perhaps a suitable $\zeta\to 0$ limit can be taken, or one may add terms that cancel the $\zeta$-dependence in the BV master equation.



\section{Regularization by smearing}

In the non-minimal pure spinor descendant field construction, the factor $(\lambda\bar\lambda)^{-1}$ appears in the descending differential operators, which has a pole at the tip of the pure spinor cone. With sufficiently many descendant fields in the integrand, one may worry about a potential divergence in the integration over the pure spinor space. On the other hand, each net factor of $(\lambda\bar\lambda)^{-1}$ is accompanied by an $r_\A$. When there are more than 11 $r$'s in the numerator, the integrand vanishes due to the pure spinor constraint relating $r_\A$ and $\bar\lambda_\A$. A priori, there could be a logarithmic divergence coming from integrating $r^{11} (\lambda\bar\lambda)^{-11}$. In the example of Born-Infeld deformation, the coefficients of such terms appear to be zero, but this isn't immediately obvious. 

It was suggested by Berkovits and Nekrasov \cite{Berkovits:2006vi} in the context of pure spinor string theory that one can regularize a potential divergence in the pure spinor integral by smearing the vertex operators in pure spinor space, in a way that preserves BRST invariance. In this section, we will adopt the same smearing operator and consider superspace Lagrangian terms built out of smeared descendant pure spinor superfields. In this way, one could eliminate potential divergences in the pure spinor space integral from the start.

A related issue is the construction of D-term deformations.
The three examples of deformed BV action of MSYM we have constructed so far are all F-term deformations. It is not clear whether D-terms can be expressed as the integral of a local expression of the superfields over the pure spinor superspace. Naively one may try to apply enough descending operators so that $r^{11}$ appears and turns the fermionic superspace integral into an integration purely over the 16 $\theta$'s. Such attempts seem to fail. In fact if we could write such an expression using the descending operators, we would also encounter a bosonic integration of $(\lambda\bar\lambda)^{-11}$ which is logarithmically divergent.
It would seem that the construction of D-terms must involve non-local terms on the pure spinor superspace,\footnote{This is not unfamiliar in the context of harmonic superspace.} where the smearing construction could be useful as well.


\subsection{A smearing operator}

First, one introduces a new bosonic pure spinor variable $f^\A$ and its fermionic counterpart $g^\A_I$, as well as their conjugate variables $\bar f_\A, \bar g_\A$, that obey the constraints\footnote{One can also generalize this construction by introducing several copies of $(f,g,\bar f,\bar g)$ variables.}
\ie
f^\A \Gamma^m_{\A\B} f^\B = f^\A\Gamma^m_{\A\B} g^\B = \bar f_\A (\Gamma^m)^{\A\B} \bar f_\B = \bar f_\A (\Gamma^m)^{\A\B} \bar g_\B = 0.
\fe
We may also identify $g^\A$ with the odd differential $df^\A$, and $\bar g_\A$ with $d\bar f_\A$. The descendant superfields generally contain terms involving some powers of $r$ and $(\lambda\bar\lambda)^{-1}$. The idea is to consider the exponential of a $Q$-exact operator that acts on the field, and effectively shifts $\lambda$ and $\bar\lambda$ by a small amount, roughly proportional to $f$ and $\bar f$, so as to smear out the pole in $(\lambda\bar\lambda)$.
The differential $Q$ will be extended to
\ie
Q = \lambda^\A d_\A + r_\A {\partial\over \partial\bar\lambda_\A} + f^\A {\partial\over \partial g^\A} + \bar g_\A {\partial\over \partial \bar f_\A}.
\fe
Note that $Q$ is well defined due to the pure spinor constraints on $r_\A, f^\A$, and $\bar g_\A$.


The smearing operator, which may also be viewed as a regulator, acts on a descendant pure spinor field $\widehat G\Psi$ as
\ie\label{sme}
[\widehat G\Psi]_\epsilon = \int e^{- \bar ff-d\bar f df} \exp(\epsilon\{Q, X\})\widehat G\Psi.
\fe
Here $X$ is a linear differential operator in the non-minimal pure spinor variables that acts on $\widehat G\Psi$, and so is $\{Q, X\}$. $\epsilon$ is a smearing parameter. In writing (\ref{sme}) we have made the identification $g=df, \bar g=d\bar f$, and the integral is understand as that of a differential top form $d^{11}fd^{11}\bar f$ over the pure spinor space of $(f,\bar f)$. Note that $\bar f f+d\bar f df = \{Q, \bar f g\}$ is $Q$-exact.

It is somewhat nontrivial to construct the desired $X$, since various pure spinor constraints must be obeyed and only certain combinations of the derivatives with respect to the pure spinor variables are allowed. The resulting expression is
\ie
X = g^\A W_\A + \bar f_\A V^\A,
\fe
where $W^\A$ and $V^\A$ are differential operators in $\lambda$ and $r$ respectively,
\ie
& W_\A = -(\lambda \bar f)^{-1} \bigg[ {1\over 4} \bar f_\A N + {1\over 8} (\Gamma^{mn} \bar f)_\A N^{mn} \bigg],
\\
& V^\A = - (f\bar\lambda)^{-1} \bigg[ {1\over 4} f^\A (\bar\lambda{\partial}_r) + {1\over 8} (\Gamma^{mn}f)^\A (\bar\lambda\Gamma_{mn} \partial_r) \bigg].
\fe
Note that $W_\A$ takes the same form as the descending operator $\hat A_\A$, except that $\bar\lambda$ has been replaced by $\bar f$. 
%
%
%
It is useful to write down the $Q$-commutator of $W_\A$ and $V^\A$, given by
%
\ie{}
& [Q, W_\A] = - d_\A - 2(\Gamma^m\lambda)_\A U_m,
\\
& U_m \equiv - {1\over 4} (\lambda\bar f)^{-1} (\bar f\Gamma_m d) + {1\over 32} (\lambda\bar f)^{-2} (\bar f\Gamma_{mnp} \bar g) N^{np}.
\fe
and
\ie
& \{Q, V^\A\} = \overline{W}^\A-{fr\over f\bar{\lambda}}V^\A,
\\
& \overline{W}^\A \equiv -(\bar \lambda f)^{-1} \bigg[ {1\over 4} f^\A \overline{N} + {1\over 8} (\Gamma^{mn} f)^\A \overline{N}^{mn} \bigg],
\\
& \overline{N} \equiv \bar\lambda\partial_{\bar\lambda} + r\partial_r,~~~~ \overline{N}_{mn} \equiv \bar\lambda\Gamma_{mn} \partial_{\bar\lambda} + r\Gamma_{mn} \partial_r.
\fe
Be cautious that $\overline{W}^\A$ is {\it not} the same as $W_\A$ simply with $\lambda,f$ and $\bar\lambda, \bar f$ exchanged, as it has the extra terms involving $r$-derivatives.


We omit lengthy algebra and record the final expression for the differential operator in the regulator exponent
\ie\label{qxone}
\{Q, X\} 
=&\widehat f \Pi_\lambda \partial_\lambda + \bar f\Pi_{\bar\lambda} \partial_{\bar\lambda} + g \Pi_\lambda d
+ \left[ \bar g \Pi_{\bar\lambda} + {1\over 16}(r\Gamma^{ijm}\bar\lambda) (\bar\lambda^{-1}\Gamma_{ij }\bar f)(\Gamma_m\bar\lambda^{-1}) \right] \partial_r,
\fe
where we used the notation
\ie
\lambda^{-1}_\A \equiv (\lambda \bar f)^{-1} \bar f_\A, ~~~~ (\bar\lambda^{-1})^\A \equiv (\bar\lambda f)^{-1} f^\A.
\fe
$\Pi_\lambda$ and $\Pi_{\bar\lambda}$ are projectors that ensures the $\lambda$ and $\bar\lambda$ derivatives are well defined. Explicitly, they are given by
\ie
& (\Pi_\lambda)_\A{}^\B\equiv \delta_\A^\B - {1\over 2}(\Gamma^m\lambda)_\A (\Gamma_m\lambda^{-1})^\B  = -{1\over 4} (\lambda^{-1})_\A \lambda^\B - {1\over 8} (\Gamma^{mn}\lambda^{-1})_\A (\lambda\Gamma_{mn})^\B,
\\
& (\Pi_{\bar\lambda})^\B{}_\A \equiv  \delta_\A^\B - {1\over 2}(\Gamma^m\bar\lambda)^\B (\Gamma_m\bar\lambda^{-1})_\A = -{1\over 4} (\bar\lambda^{-1})^\B \bar\lambda_\A - {1\over 8} (\Gamma^{mn}\bar\lambda^{-1})^\B (\bar\lambda\Gamma_{mn})_\A .
\fe
In (\ref{qxone}) we have also defined $\widehat f^\A$ as a shifted version of $f^\A$,
\ie
& \widehat f^\A \equiv f^\A - {1\over 2}(\lambda\bar f)^{-1} (g\Gamma^m\lambda)(\Gamma^m \bar g)^\A.
\fe


\subsection{Shifted pure spinor variables}

The operator $\exp(\epsilon\{Q, X\})$ acts on a field by shifting all superspace variables $x^m, \theta^\A, \lambda^\A, \bar\lambda_\A, r_\A$. First, consider the terms in $\{Q, X\}$ that involve only bosonic derivatives, dropping $g,\bar g$ dependence for the moment,
\ie\label{ggz}
\left. e^{\epsilon\{Q, X\}}\right|_{g,\bar g=0}=\exp\left[ \epsilon (\widehat f \Pi_\lambda \partial_\lambda + \bar f\Pi_{\bar\lambda} \partial_{\bar\lambda} ) \right].
\fe
The shift of $\lambda$ and $\bar\lambda$ by (\ref{ggz}) was computed by Berkovitz and Nekrasov,
\ie
&\left. e^{\epsilon\{Q, X\}}\right|_{g,\bar g=0} \lambda^\A 
= (\lambda + \epsilon f)^\A - \epsilon {(\lambda\Gamma^m f)(\Gamma_m \bar f)^\A\over 2(\lambda+\epsilon f)\bar f},
\\
&\left. e^{\epsilon\{Q, X\}}\right|_{g,\bar g=0}\bar \lambda_\A 
= (\bar \lambda + \epsilon \bar f)_\A - \epsilon {(\bar \lambda\Gamma^m \bar f)(\Gamma_m f)_\A\over 2(\bar \lambda+\epsilon \bar f)f}.
\fe
As a consistency check, note that the RHS obey the pure spinor constraint for any finite value of $\epsilon$. 

Now let us include the $g,\bar g$ dependence. The notation is simplified if we now make the identification $r_\A=d\bar\lambda_\A$, $g^\A = df^\A$, and $\bar g_\A = d\bar f_\A$.  We can write
\ie
e^{ \epsilon\{Q, X\} } F(x,\theta, \lambda, \bar\lambda, d\bar\lambda)
= F(x_\epsilon, \theta_\epsilon, \lambda_\epsilon, \bar\lambda_\epsilon, d\bar\lambda_\epsilon)
\fe
for any superfield $F$,
where
$(x_\epsilon, \theta_\epsilon, \lambda_\epsilon, \bar\lambda_\epsilon)$ are functions of $(x,\theta,\lambda,\bar\lambda)$ (independent of $d\bar\lambda$), that also depends on $f,\bar f, df,d\bar f$. It follows immediately from the structure of $\{Q,X\}$ that $r_\epsilon$ is recovered from $\lambda_\epsilon$ by differentiation with respect to $\bar\lambda$ and $\bar f$, namely
\ie
(r_\epsilon)_\A = [Q, (\bar\lambda_\epsilon)_\A] = \bar\partial (\bar\lambda_\epsilon)_\A \equiv (r\partial_{\bar\lambda}+\bar g \partial_{\bar f}) (\bar\lambda_\epsilon)_\A.
\fe
Thus, it suffices to consider the action of
\ie\label{qxsimp}
\left[ f - {1\over 2} (\lambda\bar f)^{-1} (df\Gamma^m\lambda) (d\bar f\Gamma_m) \right]\Pi_\lambda\partial_\lambda + df \Pi_\lambda \left(\partial_\theta -\Gamma^m\theta \partial_m \right) 
+ \bar f \Pi_{\bar\lambda} \partial_{\bar\lambda}  
\fe
instead of $\{Q,X\}$, on a function of $(x,\theta,\lambda,\bar\lambda)$.
The last term in (\ref{qxsimp}) commutes with the rest. And so we learn that
\ie
\bar\lambda_\epsilon = \bar\lambda + \epsilon \left[ \bar f - {\bar\lambda\Gamma^m\bar f\over 2(\bar\lambda+\epsilon \bar f) f} \Gamma_m f \right].
\fe
We do not know a simple closed formula for $x_\epsilon, \theta_\epsilon, \lambda_\epsilon$. They can be computed order by order in the fermionic variables $df, d\bar f$. 
We write below the first two terms in the expansions of $x_\epsilon, \theta_\epsilon, \lambda_\epsilon$ in $df, d\bar f$. Firstly $\lambda_\epsilon$, which is independent of $x,\theta,\bar\lambda$, takes the form
\ie
\lambda_\epsilon 
& = \lambda + \epsilon\left[ f - {\lambda\Gamma^m f\over 2 (\lambda+\epsilon f)\bar f} \Gamma_m\bar f \right] 
\\
&~~ - {\epsilon\over 2}\int_0^1 dt \left[ {df\Gamma^m\lambda \over (\lambda+t\epsilon f)\bar f } - t\epsilon { (df\Gamma^m\Gamma_n\bar f) (\lambda\Gamma^n f) \over 2((\lambda+ t\epsilon f)\bar f)^2} \right] \left[\Gamma_m d\bar f - {d\bar f(\lambda+t\epsilon f)\over \bar f (\lambda+t\epsilon f)} \Gamma_m\bar f \right] + {\cal O}(df^2 d\bar f^2).
\fe
Up to order $df^2 d\bar f^2$ terms in the expansion in $df$ and $d\bar f$, this expression is exact in $\epsilon$.
Likewise, $\theta_\epsilon$ and $x_\epsilon$ can be solved recursively,
\ie
& (\theta_\epsilon)^\A = \theta^\A + \epsilon \int_0^1 dt \, (\Pi_{\lambda_{t\epsilon}} df )^\A
\\
&~~~~~~~ = \theta^\A + \epsilon df^\A -  {\epsilon\over 2}(\Gamma^m \bar f)^\A \int_0^1 dt\, \left[ { df\Gamma_m \lambda  \over \bar f (\lambda+t\epsilon f)} - t\epsilon { (df\Gamma_m \Gamma_n\bar f) (\lambda\Gamma^n f)  \over 2(\bar f (\lambda+t\epsilon f))^2 } \right] + {\cal O}(df^2 d\bar f)  ,
\\
& x_\epsilon^m = x^m -\epsilon \int_0^1 dt \, (df \Pi_{\lambda_{t\epsilon}} \Gamma^m\theta_{t\epsilon} ).
\fe

\subsection{Superspace Lagrangian deformations using smeared fields}

A simple class of smeared deformations is the following. Suppose $S'$ is a first order deformation of the BV action, constructed out of a superspace integral of smeared descendant pure spinor superfields, typically of the form
\ie\label{smz}
S' = \int [dZ] \, {\rm Tr} \left\{ \Psi \cdots [\widehat G_i\Psi ]_{\epsilon_i}\cdots [\widehat G_{j_1}\Psi \widehat G_{j_2}\Psi ]_{\epsilon_j}\cdots \right\}
\fe
It is useful to group several descendants together, and act on with a smearing operator, in constructing a deformation that solves the BV master equation. Let us consider a total action of the form $S_2+S_3+S'$, and the BV master equation at the first order in the deformation parameter, which demands the vanishing of $(S_2, S')$ and $(S_3, S')$.\footnote{For the deformation to be nontrivial (not removable by field definition), we also need $S'$ to be not exact with respect to $(S_2,\,\cdot\,)$ and $(S_3,\,\cdot\,)$.}

Taking the BV anti-bracket $(S_2, S')$ amounts to computing the variation of $S'$ under $\delta \Psi = \eta Q\Psi$, where $\eta$ is an arbitrary odd parameter.\footnote{This variation is not to be confused with a BRST or gauge transformation; if one gauge fixes the BV action by fixing the anti-fields, then the vanishing of $(S_2, S')$ implies that the BRST transformation can be deformed in such a way that $S_2+S'$ is BRST invariant to first order in the deformation parameter of $S'$.} Consider a smeared descendant superfield that appears in the integrand of $S'$,
\ie{}
[\widehat G\Psi]_\epsilon = \int e^{-\bar f f-d\bar f\bar f} e^{\epsilon \{Q, X\}} \widehat G \Psi,
\fe
where $X$ is the first order differential operator defined as in previous subsections, and $\widehat G$ is a descending operator that involves the non-minimal variables (but only contains derivatives on $x,\theta$ and $\lambda$). 
If $\widehat G$ commutes (when it is even) or anti-commutes (when it is odd) with $Q$, we would have
\ie
\delta [\widehat G\Psi]_\epsilon =\eta Q[\widehat G\Psi]_\epsilon. 
\fe
This is the case with the non-commutative deformation and the 5-form deformation, as discussed before. Basic example of such $\widehat G$ operators are $\lambda\Gamma^m\hat\chi$ and $(\lambda\Gamma^{mn})^\A \hat F_{mn}$. It is also possible that while not all $\widehat G_i$'s commute with $Q$, a suitable linear combination of products of such descendant superfields has the desired property
\ie
\delta [\widehat G_1\Psi\widehat G_2\Psi]_\epsilon =\eta Q[\widehat G_1\Psi \widehat G_2\Psi\cdots]_\epsilon.
\fe
We have seen this in the example of the Born-Infeld deformation, in the combination $(\lambda\Gamma^m\Psi)(\lambda\Gamma^n\Psi)(\hat F_{mn}\Psi)$. If all smeared factors in (\ref{smz}) have this property, then $S'$ obeys $(S_2, S')=0$. On the other hand, it is easy to see by similar arguments that the $\epsilon$-dependence is $S_2$-exact, which means that the deformation by smearing is independent of $\epsilon$, at least when $\epsilon$ is nonzero.

In the non-Abelian MSYM theory, we also need to demand the vanishing of $(S_3, S')$, which is equivalent to the invariance of $S'$ under $\delta\Psi = \eta \Psi^2$. This is the translation-invariant cyclic cocycle condition as discussed before. It seems difficult to satisfy this cocycle condition with the product of generic smeared superfields. On the other hand, the cocycle condition can be satisfied if we take $\epsilon\to 0$ limit on $[\widehat G\Psi\cdots]_\epsilon$. Note that when the naive product of such field operators vanishes due to more than 11 powers of $r$'s, the smeared product can potentially be nontrivial in the $\epsilon\to 0$ limit (after the pure spinor superspace integral).

We have seen that in the descending operators $\hat\chi^\A, \hat F_{mn}$, etc., each pole factor $(\lambda\bar\lambda)^{-1}$ is accompanied by a factor of $r_\A$. Whenever there is potentially an $n$-th order divergence coming from integrating $(\lambda\bar\lambda)^{-11-n}$ over the pure spinor space, we also have a factor formally of the form $r^{11+n}$ in the numerator that vanishes. After replacing some of the descendant superfields by their smeared versions, some of the $r$'s will be shifted to $r_\epsilon = \bar\partial\bar\lambda_\epsilon$, so that the numerator is no longer identically zero, but of order $\epsilon^n$. In the denominator, some of $(\lambda\bar\lambda)$'s will be replaced by $(\lambda_\epsilon \bar\lambda_\epsilon)$, and typically the divergent $(\lambda,\bar\lambda)$-integral will be of order $\epsilon^{-n}$. After this ``regularization", the resulting functional can stay finite if we take $\epsilon\to 0$ in the end.\footnote{One might worry about the terms that involve $((\lambda+\epsilon f)\bar f)^{-1}$ or $((\bar\lambda+\epsilon\bar f) f)^{-1}$ in the formula for the shifted pure spinor variables giving rise to extra poles in $\epsilon$. A more careful inspection of the $\lambda_\epsilon\bar\lambda_\epsilon$ factors in the denominator shows that this doesn't happen. }

It is clear that the new terms in the integrand that arise this way in the $\epsilon\to 0$ limit will always contain $r^{11}$, which then absorbs the Grassmannian $r$-integral, leaving no room for a $\theta$-dependent factor from the regulator $e^{-\zeta(\lambda\bar\lambda + r\theta)}$. The result then looks like an integral of descendant superfields over the full $\theta$-superspace. These appear to be D-terms. 
We don't yet have a proposal for the construction of the general D-terms, which we leave for future work.

\section{Discussion}

The main result of this paper is a construction of an all-order Born-Infeld deformation of the MSYM theory, in the non-minimal pure spinor superspace formalism. It would be nice to produce the corresponding all-order deformed superfield equation of motion in the ordinary superfield $A_\A(x,\theta)$, after eliminating the auxiliary fields having to do with the non-minimal variables. In practice, as explained in section \ref{recmin}, this amounts to finding the minimal representatives of certain non-minimal pure spinor cohomology classes.

An unsatisfying aspect of the story is that we don't know how to write the general D-terms in the non-minimal superspace formalism (which one might have expected to be the easiest thing). This question is also related to how to write the D-term deformation of the equation of motion in terms of the on-shell superfield $A_\A(x,\theta)$. The answer to the latter question is nontrivial though in principle known: as explained by \cite{Movshev:2005ei, Movshev:2009ba} and also discussed in \cite{partone}, a gauge invariant expression ${\rm tr}(G)$ in component fields is mapped to a deformation of the superfield equation by the composition of the Connes differential with a map $\delta$ that amounts to performing a full superspace integral, but is constructed rather inexplicitly through a spectral sequence argument that involves lifting the relevant chain complex to a complex of vector bundles over the projective pure spinor space.

We suspect that the D-terms must be written as a non-local expression in pure spinor superspace. This is presumably closely related to the regularization of \cite{Berkovits:2006vi}, which is relevant in computing the D-term contributions in higher genus string amplitudes. Though we have constructed {\it an} all-order Born-Infeld deformation, in principle it may differ from {\it the} Born-Infeld theory that arises as the $\A'$-expansion of the low energy effective theory of open strings on D-branes, by some D-term ambiguity. A potential application of our construction of the all-order Born-Infeld action, as well as a test of its relation to the open string effective action, would be to find some nontrivial nonlinear solutions to the equation of motion in the non-minimal pure spinor superfields and compare it with D-brane configurations (along the lines of \cite{Constable:1999ac}). It would also be interesting to directly connect our construction to open string disc amplitudes in the pure spinor formalism.

Ultimately, the non-minimal pure spinor formalism for constructing higher derivative terms may be most useful in maximally supersymmetric supergravity theories. In \cite{Cederwall:2009ez,Cederwall:2010tn} Cederwall wrote down a remarkable manifestly supersymmetric complete BV action for 11-dimensional supergravity in pure spinor superspace. It would be interesting to construct the $R^4$ deformation in this formalism.



\bigskip

\section*{Acknowledgments}

We are grateful to Shu-Heng Shao for collaboration at the initial stage of the project, and to Clay Cordova, Thomas Dumitrescu, Ken Intriligator, Daniel Jafferis, and Nati Seiberg for helpful discussions. We would like to thank the organizers of the workshop {\it String Geometry and Beyond} at Soltis Center, Costa Rica, the KITP program {\it New Methods in Nonperturbative Quantum Field Theory}, and especially the support of KITP during the course of this work. This work is supported in part by a KITP Graduate Fellowship, a Sloan Fellowship, a Simons Investigator Award from the Simons Foundation,  NSF Award PHY-0847457, and by the Fundamental Laws Initiative 
Fund at Harvard University.

\appendix

\section{Siegel gauge and the $b$ ghost}

In order to go from the BV action functional to a gauge fixed BRST invariant action, a gauge fixing condition must be imposed that determines the anti-fields in terms of the ordinary gauge fields and the ghosts. Note that the  gauge fixing procedure in the BV formalism is different from that of an ordinary gauge invariant classical action, in that one should impose the gauge fixing condition {\it before}  applying the variational principle on the action functional to obtain the equation of motion. In the pure spinor superspace formulation of the BV action of MSYM, it is a priori not clear how to separate $\Psi(x,\theta,\lambda,\bar\lambda, r)$ into ordinary gauge fields and anti-fields. It has been suggested that an appropriate gauge fixing condition is the Siegel gauge \cite{Aisaka:2009yp,Cederwall:2011vy, Cederwall:2013vba}
\ie
b \Psi = 0,
\fe
where $b$ is a second order differential operator that obeys
\ie
\{Q, b\} = \partial^m \partial_m.
\fe
The $b$ ghost admits the following representation\footnote{The signs in our formula differ slightly from those of \cite{Cederwall:2013vba}.}
\ie
b &= -{1\over 2}(\lambda\bar\lambda)^{-1} (\bar\lambda\Gamma^m d) \partial_m + {1\over 16}(\lambda\bar\lambda)^{-2}(\bar\lambda\Gamma^{mnp} r) \left(N_{mn} \partial_p - {1\over 24} d\Gamma_{mnp} d\right)
\\
& ~~~ + {1\over 64} (\lambda\bar\lambda)^{-3} (r\Gamma^{mnp} r) (\bar\lambda\Gamma_m d)N_{np} 
- {1\over 1024} (\lambda\bar\lambda)^{-4} (\bar\lambda\Gamma^{mns} r) (r \Gamma^{pq}{}_s r) N_{mn} N_{pq}.
\fe
This expression can be expressed simply in terms of the descending operators
\ie
b = \partial_m \hat A^m - {1\over 2} d_\A \hat \chi^\A + {1\over 4} N_{mn} \hat F^{mn}.
\fe
This is reminiscent of the form of the integrated massless vertex operator in pure spinor string theory.
Indeed it is easy to verify
\ie
\{Q, b\} &= \partial_m \{Q,\hat A^m\} + \partial_m(\lambda\Gamma^m\hat\chi^\A) + {1\over 2}d_\A [Q, \hat\chi^\A]
- {1\over 4} (\lambda\Gamma_{mn} d)\hat F^{mn} + {1\over 4} N_{mn}\{Q, \hat F^{mn}\}
\\
&= \partial_m \partial^m+ {1\over 2} N^{mn}\lambda\Gamma_m\hat \eta_n = \partial_m\partial^m.
\fe
Another property of the $b$ ghost operator is $b^2=0$. This is necessary for the Siegel gauge condition to be compatible with BV master equation. 

After fixing to Siegel gauge, the equation of motion may be obtained from the BV action of the form $S_2+S_{int}$ as
\ie
Q \Psi + {\delta S_{int}\over \delta\Psi} + b\Lambda = 0,
\fe
where $\Lambda$ is an arbitrary Lagrangian multiplier superfield. Acting on this equation with $b$, using the Siegel gauge condition and the nilpotency of $b$, we obtain
\ie
\Box \Psi + b {\delta S_{int}\over \delta\Psi} = 0.
\fe


Let us inspect the Siegel gauge condition more explicitly in the simple example of free Abelian theory. Consider a solution to $Q\Psi=0$ that involves only the minimal pure spinor variables of the form
\ie
\Psi(x,\theta,\lambda) = (\lambda\Gamma^m\theta) a_m(x) + {1\over 4}(\lambda\Gamma_m\theta) (\theta\Gamma^{mnp}\theta) \partial_n a_p+ \cdots.
\fe
Such a $\Psi$ does not obey Siegel gauge condition, since
\ie
b\Psi &= -{1\over 2}(\lambda\bar\lambda)^{-1} (\bar\lambda\Gamma^m\Gamma^n\lambda) \partial_m a_n(x) + {\cal O}(\theta)
\\
&=-{1\over 2}\partial_m a^m(x) - {1\over 2} (\lambda\bar\lambda)^{-1} (\bar\lambda\Gamma^{mn}\lambda) \partial_m a_n(x) + {\cal O}(\theta).
\fe 
While we can set $\partial_m a^m$ to zero by imposing Lorentz gauge condition on $a_m$, $\partial_m a_n$ is a nontrivial field strength and cannot be removed this way. We would like to add to $\Psi$ some $Q$-exact terms to go to Siegel gauge. Using the non-minimal variables, we can write $(\lambda\Gamma^m\theta) a_m+\cdots$ as an exact expression with respect to $\lambda^\A d_\A$ (which is not the same as $Q$ in the non-minimal formalism)
\ie
\Psi(x,\theta,\lambda) = (\lambda^\A d_\A) \left[ {1\over 8}(\lambda\bar\lambda)^{-1}(\bar\lambda\Gamma_{np}\lambda) (\theta\Gamma^{mnp}\theta)  a_m \right] + \cdots,
\fe
and now remove the term $(\lambda\Gamma^m\theta) a_m$ by shifting $\Psi$ to
\ie
& \Psi' = \Psi - Q\left[ {1\over 8}(\lambda\bar\lambda)^{-1} (\bar\lambda\Gamma_{np}\lambda)(\theta\Gamma^{mnp}\theta)  a_m + \cdots \right] 
\\
&=  -{1\over 8}(\lambda\bar\lambda)^{-1} (r\Gamma_{np}\lambda)(\theta\Gamma^{mnp}\theta)  a_m
+ {1\over 8}(\lambda\bar\lambda)^{-2} (r\lambda) (\bar\lambda\Gamma_{np}\lambda) (\theta\Gamma^{mnp}\theta)  a_m+\cdots
\fe
The physical degree of freedom $a_m$ is now moved to the $r\theta^2$ component of $\Psi'$. By repeating such a procedure we should be able to put the shifted $\Psi$ in Siegel gauge. In the end, $a_m(x)$ will no longer sit in the $r^0$ component of $\Psi$.

\bibliography{offshellrefs} 

\providecommand{\href}[2]{#2}\begingroup\raggedright\begin{thebibliography}{10}

\bibitem{Siegel:1978yi}
W.~Siegel, {\it {Superfields in Higher Dimensional Space-time}},  {\em
  Phys.Lett.} {\bf B80} (1979) 220.

\bibitem{Movshev:2003ib}
M.~Movshev and A.~S. Schwarz, {\it {On maximally supersymmetric Yang-Mills
  theories}},  {\em Nucl.Phys.} {\bf B681} (2004) 324--350,
  [\href{http://xxx.lanl.gov/abs/hep-th/0311132}{{\tt hep-th/0311132}}].

\bibitem{Movshev:2004aw}
M.~Movshev and A.~S. Schwarz, {\it {Algebraic structure of Yang-Mills theory}},
   \href{http://xxx.lanl.gov/abs/hep-th/0404183}{{\tt hep-th/0404183}}.

\bibitem{Movshev:2005ei}
M.~Movshev, {\it {Deformation of maximally supersymmetric Yang-Mills theory in
  dimensions 10. An Algebraic approach}},
  \href{http://xxx.lanl.gov/abs/hep-th/0601010}{{\tt hep-th/0601010}}.

\bibitem{Movshev:2009ba}
M.~Movshev and A.~Schwarz, {\it {Supersymmetric Deformations of Maximally
  Supersymmetric Gauge Theories}},  {\em JHEP} {\bf 1209} (2012) 136,
  [\href{http://xxx.lanl.gov/abs/0910.0620}{{\tt arXiv:0910.0620}}].

\bibitem{Cederwall:2011vy}
M.~Cederwall and A.~Karlsson, {\it {Pure spinor superfields and Born-Infeld
  theory}},  {\em JHEP} {\bf 1111} (2011) 134,
  [\href{http://xxx.lanl.gov/abs/1109.0809}{{\tt arXiv:1109.0809}}].

\bibitem{Cederwall:2013vba}
M.~Cederwall, {\it {Pure spinor superfields -- an overview}},
  \href{http://xxx.lanl.gov/abs/1307.1762}{{\tt arXiv:1307.1762}}.

\bibitem{Berkovits:2001rb}
N.~Berkovits, {\it {Covariant quantization of the superparticle using pure
  spinors}},  {\em JHEP} {\bf 0109} (2001) 016,
  [\href{http://xxx.lanl.gov/abs/hep-th/0105050}{{\tt hep-th/0105050}}].

\bibitem{partone}
C.-M. Chang, Y.-H. Lin, Y.~Wang, and X.~Yin, {\it {Deformations with Maximal
  Supersymmetries, Part 1: On-shell Formulation}},
  \href{http://xxx.lanl.gov/abs/1403.0545}{{\tt arXiv:1403.0545}}.

\bibitem{Nilsson:1985cm}
B.~Nilsson, {\it {Pure Spinors as Auxiliary Fields in the Ten-dimensional
  Supersymmetric {Yang-Mills} Theory}},  {\em Class.Quant.Grav.} {\bf 3} (1986)
  L41.

\bibitem{Tonin:1991ii}
M.~Tonin, {\it {World sheet supersymmetric formulations of Green-Schwarz
  superstrings}},  {\em Phys.Lett.} {\bf B266} (1991) 312--316.

\bibitem{Howe:1991mf}
P.~S. Howe, {\it {Pure spinors lines in superspace and ten-dimensional
  supersymmetric theories}},  {\em Phys.Lett.} {\bf B258} (1991) 141--144.

\bibitem{Howe:1991bx}
P.~S. Howe, {\it {Pure spinors, function superspaces and supergravity theories
  in ten-dimensions and eleven-dimensions}},  {\em Phys.Lett.} {\bf B273}
  (1991) 90--94.

\bibitem{Berkovits:2000fe}
N.~Berkovits, {\it {Super Poincare covariant quantization of the superstring}},
   {\em JHEP} {\bf 0004} (2000) 018,
  [\href{http://xxx.lanl.gov/abs/hep-th/0001035}{{\tt hep-th/0001035}}].

\bibitem{Berkovits:2000nn}
N.~Berkovits, {\it {Cohomology in the pure spinor formalism for the
  superstring}},  {\em JHEP} {\bf 0009} (2000) 046,
  [\href{http://xxx.lanl.gov/abs/hep-th/0006003}{{\tt hep-th/0006003}}].

\bibitem{Berkovits:2000ph}
N.~Berkovits and B.~C. Vallilo, {\it {Consistency of superPoincare covariant
  superstring tree amplitudes}},  {\em JHEP} {\bf 0007} (2000) 015,
  [\href{http://xxx.lanl.gov/abs/hep-th/0004171}{{\tt hep-th/0004171}}].

\bibitem{Berkovits:2001ue}
N.~Berkovits and P.~S. Howe, {\it {Ten-dimensional supergravity constraints
  from the pure spinor formalism for the superstring}},  {\em Nucl.Phys.} {\bf
  B635} (2002) 75--105, [\href{http://xxx.lanl.gov/abs/hep-th/0112160}{{\tt
  hep-th/0112160}}].

\bibitem{Berkovits:2001us}
N.~Berkovits, {\it {Relating the RNS and pure spinor formalisms for the
  superstring}},  {\em JHEP} {\bf 0108} (2001) 026,
  [\href{http://xxx.lanl.gov/abs/hep-th/0104247}{{\tt hep-th/0104247}}].

\bibitem{Matone:2002ft}
M.~Matone, L.~Mazzucato, I.~Oda, D.~Sorokin, and M.~Tonin, {\it {The
  Superembedding origin of the Berkovits pure spinor covariant quantization of
  superstrings}},  {\em Nucl.Phys.} {\bf B639} (2002) 182--202,
  [\href{http://xxx.lanl.gov/abs/hep-th/0206104}{{\tt hep-th/0206104}}].

\bibitem{Berkovits:2002zk}
N.~Berkovits, {\it {ICTP lectures on covariant quantization of the
  superstring}},  \href{http://xxx.lanl.gov/abs/hep-th/0209059}{{\tt
  hep-th/0209059}}.

\bibitem{Berkovits:2004px}
N.~Berkovits, {\it {Multiloop amplitudes and vanishing theorems using the pure
  spinor formalism for the superstring}},  {\em JHEP} {\bf 0409} (2004) 047,
  [\href{http://xxx.lanl.gov/abs/hep-th/0406055}{{\tt hep-th/0406055}}].

\bibitem{Berkovits:2004tw}
N.~Berkovits and D.~Z. Marchioro, {\it {Relating the Green-Schwarz and pure
  spinor formalisms for the superstring}},  {\em JHEP} {\bf 0501} (2005) 018,
  [\href{http://xxx.lanl.gov/abs/hep-th/0412198}{{\tt hep-th/0412198}}].

\bibitem{Berkovits:2005bt}
N.~Berkovits, {\it {Pure spinor formalism as an N=2 topological string}},  {\em
  JHEP} {\bf 0510} (2005) 089,
  [\href{http://xxx.lanl.gov/abs/hep-th/0509120}{{\tt hep-th/0509120}}].

\bibitem{Berkovits:2006vi}
N.~Berkovits and N.~Nekrasov, {\it {Multiloop superstring amplitudes from
  non-minimal pure spinor formalism}},  {\em JHEP} {\bf 0612} (2006) 029,
  [\href{http://xxx.lanl.gov/abs/hep-th/0609012}{{\tt hep-th/0609012}}].

\bibitem{Witten:1985cc}
E.~Witten, {\it {Noncommutative Geometry and String Field Theory}},  {\em
  Nucl.Phys.} {\bf B268} (1986) 253.

\bibitem{Bergshoeff:1986jm}
E.~Bergshoeff, M.~Rakowski, and E.~Sezgin, {\it {HIGHER DERIVATIVE
  SUPERYANG-MILLS THEORIES}},  {\em Phys.Lett.} {\bf B185} (1987) 371.

\bibitem{Bergshoeff:2000cx}
E.~Bergshoeff, M.~de~Roo, and A.~Sevrin, {\it {On the supersymmetric nonAbelian
  Born-Infeld action}},  {\em Fortsch.Phys.} {\bf 49} (2001) 433--440,
  [\href{http://xxx.lanl.gov/abs/hep-th/0011264}{{\tt hep-th/0011264}}].

\bibitem{Bergshoeff:2001dc}
E.~Bergshoeff, A.~Bilal, M.~de~Roo, and A.~Sevrin, {\it {Supersymmetric
  nonAbelian Born-Infeld revisited}},  {\em JHEP} {\bf 0107} (2001) 029,
  [\href{http://xxx.lanl.gov/abs/hep-th/0105274}{{\tt hep-th/0105274}}].

\bibitem{Sevrin:2001ha}
A.~Sevrin, J.~Troost, and W.~Troost, {\it {The nonAbelian Born-Infeld action at
  order F**6}},  {\em Nucl.Phys.} {\bf B603} (2001) 389--412,
  [\href{http://xxx.lanl.gov/abs/hep-th/0101192}{{\tt hep-th/0101192}}].

\bibitem{Cederwall:2001bt}
M.~Cederwall, B.~E. Nilsson, and D.~Tsimpis, {\it {The Structure of maximally
  supersymmetric Yang-Mills theory: Constraining higher order corrections}},
  {\em JHEP} {\bf 0106} (2001) 034,
  [\href{http://xxx.lanl.gov/abs/hep-th/0102009}{{\tt hep-th/0102009}}].

\bibitem{Cederwall:2001td}
M.~Cederwall, B.~E. Nilsson, and D.~Tsimpis, {\it {D = 10 superYang-Mills at
  O(alpha-prime**2)}},  {\em JHEP} {\bf 0107} (2001) 042,
  [\href{http://xxx.lanl.gov/abs/hep-th/0104236}{{\tt hep-th/0104236}}].

\bibitem{Cederwall:2001dx}
M.~Cederwall, B.~E. Nilsson, and D.~Tsimpis, {\it {Spinorial cohomology and
  maximally supersymmetric theories}},  {\em JHEP} {\bf 0202} (2002) 009,
  [\href{http://xxx.lanl.gov/abs/hep-th/0110069}{{\tt hep-th/0110069}}].

\bibitem{Collinucci:2002ac}
A.~Collinucci, M.~De~Roo, and M.~Eenink, {\it {Supersymmetric Yang-Mills theory
  at order alpha-prime**3}},  {\em JHEP} {\bf 0206} (2002) 024,
  [\href{http://xxx.lanl.gov/abs/hep-th/0205150}{{\tt hep-th/0205150}}].

\bibitem{Howe:2010nu}
P.~Howe, U.~Lindstrom, and L.~Wulff, {\it {D=10 supersymmetric Yang-Mills
  theory at $\alpha'^4$}},  {\em JHEP} {\bf 1007} (2010) 028,
  [\href{http://xxx.lanl.gov/abs/1004.3466}{{\tt arXiv:1004.3466}}].

\bibitem{Bossard:2010pk}
G.~Bossard, P.~Howe, U.~Lindstrom, K.~Stelle, and L.~Wulff, {\it {Integral
  invariants in maximally supersymmetric Yang-Mills theories}},  {\em JHEP}
  {\bf 1105} (2011) 021, [\href{http://xxx.lanl.gov/abs/1012.3142}{{\tt
  arXiv:1012.3142}}].

\bibitem{Batalin:1981jr}
I.~Batalin and G.~Vilkovisky, {\it {Gauge Algebra and Quantization}},  {\em
  Phys.Lett.} {\bf B102} (1981) 27--31.

\bibitem{Fuster:2005eg}
A.~Fuster, M.~Henneaux, and A.~Maas, {\it {BRST quantization: A Short review}},
   {\em Int.J.Geom.Meth.Mod.Phys.} {\bf 2} (2005) 939--964,
  [\href{http://xxx.lanl.gov/abs/hep-th/0506098}{{\tt hep-th/0506098}}].

\bibitem{Ishibashi:1996xs}
N.~Ishibashi, H.~Kawai, Y.~Kitazawa, and A.~Tsuchiya, {\it {A Large N reduced
  model as superstring}},  {\em Nucl.Phys.} {\bf B498} (1997) 467--491,
  [\href{http://xxx.lanl.gov/abs/hep-th/9612115}{{\tt hep-th/9612115}}].

\bibitem{galperin2007harmonic}
A.~Galperin, E.~Ivanov, V.~Ogievetsky, and E.~Sokatchev, {\em Harmonic
  Superspace}.
\newblock Cambridge Monographs on Mathematical Physics. Cambridge University
  Press, 2007.

\bibitem{Cederwall:2009ez}
M.~Cederwall, {\it {Towards a manifestly supersymmetric action for
  11-dimensional supergravity}},  {\em JHEP} {\bf 1001} (2010) 117,
  [\href{http://xxx.lanl.gov/abs/0912.1814}{{\tt arXiv:0912.1814}}].

\bibitem{lie}
M.~A.~A. van Leeuwen, A.~M. Cohen, and B.~Lisser, {\it Lie, a package for lie
  group computations},  {\em Computer Algebra Nederland} (1992).

\bibitem{Aganagic:1996nn}
M.~Aganagic, C.~Popescu, and J.~H. Schwarz, {\it {Gauge invariant and gauge
  fixed D-brane actions}},  {\em Nucl.Phys.} {\bf B495} (1997) 99--126,
  [\href{http://xxx.lanl.gov/abs/hep-th/9612080}{{\tt hep-th/9612080}}].

\bibitem{LodayQuillen}
J.~Loday and D.~Quillen, {\it {Cyclic Homology and the Lie Algebra Homology of
  Matrices}},  {\em Comment. Math. Holy.} {\bf 59} (1984) 565.

\bibitem{Chang:2013fba}
C.-M. Chang and X.~Yin, {\it {1/16 BPS States in N=4 SYM}},  {\em Phys.Rev.}
  {\bf D88} (2013) 106005, [\href{http://xxx.lanl.gov/abs/1305.6314}{{\tt
  arXiv:1305.6314}}].

\bibitem{2002math......7118K}
M.~{Khalkhali} and B.~{Rangipour}, {\it {Invariant Cyclic Homology}},  {\em
  ArXiv Mathematics e-prints} (July, 2002)
  [\href{http://xxx.lanl.gov/abs/math/0207118}{{\tt math/0207118}}].

\bibitem{Seiberg:1999vs}
N.~Seiberg and E.~Witten, {\it {String theory and noncommutative geometry}},
  {\em JHEP} {\bf 9909} (1999) 032,
  [\href{http://xxx.lanl.gov/abs/hep-th/9908142}{{\tt hep-th/9908142}}].

\bibitem{Constable:1999ac}
N.~R. Constable, R.~C. Myers, and O.~Tafjord, {\it {The Noncommutative bion
  core}},  {\em Phys.Rev.} {\bf D61} (2000) 106009,
  [\href{http://xxx.lanl.gov/abs/hep-th/9911136}{{\tt hep-th/9911136}}].

\bibitem{Cederwall:2010tn}
M.~Cederwall, {\it {D=11 supergravity with manifest supersymmetry}},  {\em
  Mod.Phys.Lett.} {\bf A25} (2010) 3201--3212,
  [\href{http://xxx.lanl.gov/abs/1001.0112}{{\tt arXiv:1001.0112}}].

\bibitem{Aisaka:2009yp}
Y.~Aisaka and N.~Berkovits, {\it {Pure Spinor Vertex Operators in Siegel Gauge
  and Loop Amplitude Regularization}},  {\em JHEP} {\bf 0907} (2009) 062,
  [\href{http://xxx.lanl.gov/abs/0903.3443}{{\tt arXiv:0903.3443}}].

\end{thebibliography}\endgroup
\bibliographystyle{JHEP}
 
\end{document}